\documentclass{jaa}
\usepackage{natbib}

\usepackage{gensymb} 
\usepackage{graphicx}	
\usepackage{amsmath}	
\usepackage{mathtools}
\usepackage{hyperref}
\usepackage[mathscr]{euscript}
\usepackage{nccmath}
\DeclareUnicodeCharacter{2212}{-}

\begin{document}\sloppy

\title{GAMA-23 deep field with uGMRT Band-2: source catalogue and source count}


\author{RASHMI SAGAR\textsuperscript{1,*}, ABHIRUP DATTA\textsuperscript{1} and AISHRILA MAZUMDER\textsuperscript{2}}
\affilOne{\textsuperscript{1}Department of Astronomy, Astrophysics and Space Engineering, Indian Institute of Technology, Indore 453552, India.\\}
\affilTwo{\textsuperscript{2}Jodrell Bank Centre for Astrophysics, Department of Physics and Astronomy, The University of Manchester, Manchester M13 9PL, UK.}


\twocolumn[{

\maketitle

\corres{rashmisagar777@gmail.com}

\msinfo{31 May 2026}{31 May 2026}

\begin{abstract}
We present the low-frequency GAMA-23 field of the Galaxy and Mass Assembly (GAMA) survey with upgraded Giant Meterwave Radio Telescope (uGMRT) observations in Band-2 ($120-250$\,MHz). The on-source observation was conducted for ${\sim}5$ hours. The final image achieved an off-source RMS noise of 320\,$\mu$Jy beam$^{-1}$, covering an area of $5.86\,\text{deg}^2$ with restoring beam size of $15.4''{\times}8.2''$ at the central frequency of 183\,MHz. The catalogue derived from the radio image includes 569 sources with a flux density limit of ${\geq}\,5{\sigma}$. The flux density measurements and the positional accuracy are cross-verified and compared with other radio catalogues available in the literature. The spectral indices derived from fitting a power law are -0.68$^{+0.18}_{-0.23}$ for the ASKAP EMU catalogue and  -0.72$^{+0.19}_{-0.24}$ for the NVSS catalogue. The normalised source count derived from this catalogue is consistent with models and observations with a nominal catalogue threshold of 1.7\,mJy, while the sensitivity varies across the image. 
\end{abstract}


}]


\doinum{12.3456/s78910-011-012-3}
\artcitid{\#\#\#\#}
\volnum{000}
\year{0000}
\pgrange{1--}
\setcounter{page}{1}
\lp{1}

\section{Introduction}

Deep radio continuum surveys below 1\,GHz provide a powerful window into the dust-free universe. In deep extragalactic fields, synchrotron emission is the dominant radio mechanism arising from different source populations such as star-forming galaxies (SFG), radio galaxies and diffuse extended sources \citep{1992ARA&A..30..575C, 2004MNRAS.355.1053D, 2016MNRAS.457..730P}. Within the Galaxy, emission arises from supernova remnants \citep{2017A&A...605A..58A} and from diffuse synchrotron emission in the Galactic plane \citep{1982A&AS...47....1H, 1999A&A...345..380S}. Recent low-frequency surveys provide information about these source populations, including active galactic nuclei (AGN), as their feedback mechanisms drive galaxy evolution and affect the intergalactic medium \citep{best2005host, 2012ARA&A..50..455F, 2020A&A...641A..85S}. \\
The LOFAR Two-meter Sky Survey \citep[LoTSS,][]{2022A&A...659A...1S} with LoTSS deep field \citep{2025A&A...695A..80S} at 150\,MHz from the Low Frequency Array \citep[LOFAR,][]{2013A&A...556A...2V} covers the deepest survey of Northern sky, while TIFR GMRT Sky Survey - Alternate Data Release \citep[TGSS-ADR1,][]{2017A&A...598A..78I} at 150\,MHz from the Giant Meterwave Radio Telescope \citep[GMRT,][]{2017A&A...598A..78I}, GaLactic and Extragalactic All-sky Murchison Widefield Array (GLEAM)-X survey \citep{2022PASA...39...35H} at $72-231$\,MHz from Murchison Wide-field Array \citep[MWA,][]{2013PASA...30...31B}, Sydney University Molonglo Sky Survey \citep[SUMSS,][]{2003MNRAS.342.1117M} at 843\,MHz, Evolutionary Map of the Universe \citep[EMU,][]{2021PASA...38...46N, 2025PASA...42...71H} survey at 887.5\,MHz and the Rapid ASKAP Continuum Survey \citep[RACS,][]{2021PASA...38...58H, 2023PASA...40...34D} from the Australian Square Kilometre Array Pathfinder\citep[ASKAP,][]{2014PASA...31...41H}, NRAO VLA Sky Survey \citep[NVSS,][]{1998AJ....115.1693C} at 1400\,MHz from Very Large Array \citep[VLA,][]{2011ApJ...739L...1P} cover both the Northern and Southern sky. These surveys show that the radio source population is dominated by synchrotron emitting AGNs and steep-spectrum sources, while SFGs and extended emission become significant in deeper observations \citep{2021A&A...648A...5M, 2021PASA...38...58H, 2022PASA...39...35H}. Subsequently, the differential source count becomes important for understanding galaxy evolution through the flux densities of the source population \citep{2014MNRAS.440.2791V, 2016A&ARv..24...13P, 2017A&A...602A...1S, 2023MNRAS.523.1729B}. It has been observed that AGNs (especially radio-loud AGNs) and radio galaxies dominate at the high end of the flux scale, and SFGs dominate at the faint end \citep{2009ApJ...694..235P, 2015MNRAS.452.1263P}. Thus, the source counts trend shows flattening at sub-mJy levels \citep{2019MNRAS.490..243C, 2020MNRAS.491.1127O}. However, it is found that radio-quiet AGNs also contribute at the lower flux density end though their origin is widely disputed \citep{2013MNRAS.436.3759B, 2017A&A...602A...2S}.\\
The Galaxy and Mass Assembly Project\footnote{\url{http://www.gama-survey.org/}} \citep[GAMA,][]{2016MNRAS.455.3911D} is multiwavelength survey targeting five fields $-$ G02, G09, G12, G15, and G23, in the Southern sky and covers area of ${\sim}286\,\text{deg}^2$. One of the fields observed is the GAMA-23 field with coordinates 22h36m00s $\leq \alpha_{J2000} \leq$ 23h24m00s and -30$^\circ$ $\leq \delta_{J2000} \leq$ -35$^\circ$. The GAMA-23 field lies in a well-studied multiwavelength region with extensive optical, infrared, and spectroscopic coverage. In previous studies, the GAMA-23 field is covered by the TGSS 150\,MHz survey. Recent deep-field surveys include the ASKAP EMU survey \citep{2022MNRAS.512.6104G} and the RACS survey \citep{2021PASA...38...58H} at 887.5\,MHz. The ASKAP GAMA-23 field survey covers ${\sim}\,83\,\text{deg}^2$ area with resolution of $10''$ and the central RMS noise 38\,$\mu$Jy beam$^{-1}$. We have also studied recently GAMA-23 field with uGMRT at 325\,MHz with RMS noise $109\,\mu$Jy beam$^{-1}$ over the area of ${\sim}\,65\,\text{deg}^2$ \citep{2026AJ....172..197S}. At the MHz regime, low-frequency observations are highly affected by time-varying ionospheric fluctuations, frequency-dependent primary beam, RFI, and instrument systematics, which can introduce amplitude and phase errors and therefore limit the image quality. However, adapting direction-dependent calibration, a precisely modelled primary beam, and instrument-specific improved techniques to handle RFI are carving the way out. We are introducing deeper low-frequency observation with the upgraded GMRT \citep[uGMRT,][]{2017CSci..113..707G} wideband Band-2 ($120-250$\,MHz) to explore the deep extragalactic field GAMA-23. This frequency-wideband opens the window onto the source population with higher brightness fluctuations, dominant synchrotron emission, and steep-spectrum sources. When combined with $<1$\,GHz radio catalogues, Band-2 data is useful to derive spectral indices and to understand the physical nature of identified sources in the field. \\
This work observed the GAMA-23 field in the frequency range of 120\,-\,250\,MHz with the GMRT wide-band Backend \citep[GWB,][]{2017JAI.....641011R}. This paper will complement the ASKAP EMU survey at 887.5\,MHz, the GLEAM-X survey at 200\,MHz and the uGMRT TGSS survey at 150\,MHz. This work covers a subset of both the TGSS 150\,MHz survey and the uGMRT 325\,MHz observations, significantly deeper observations. It will also provide enhanced information on the emission at 183\,MHz from radio sources, contributing to our understanding of their fundamental physics. Our primary emphasis in this paper is on data reduction, deep-field imaging, and the source catalogue. In the subsequent paper, we will investigate the cosmological aspects of the continuum emission using the angular power spectrum (APS) approach on Band-2 wideband and Band-3 data, aiming to investigate the characterisation of the Diffuse Galactic Synchrotron Emission (DGSE) in detail.\\
The paper discussed the details of the uGMRT observation, data reduction and imaging in Section~\ref{sec:observation}. Section~\ref{sec:catalogue} describes the product catalogue from the final image and discusses the comparison of our catalogue with other previous surveys and catalogues. Section~\ref{sec:count} estimates the false detection rate and incompleteness for the catalogue and applies them to the raw source count to derive the normalised corrected source count. Section~\ref{sec:conclusion} summarises the results of this work and prospects.   

\section{Observation and Data Analysis}\label{sec:observation}

\subsection{uGMRT Band-2 data}\label{sec:data}

\begin{table}[htb]
\tabularfont
\caption{Details of the uGMRT observation of GAMA-23 field at Band-2 ($120-250$\,MHz).}\label{table:tab1} 
\begin{tabular}{ll}
\topline
Project code & 42\_034\\\midline
Observation date & 18 Jul 2022 \\
Total observation time & 09 hours \\
Frequency range & $120-250$\,MHz \\
Integration time & 5.37 sec \\
Number of channels & 8192 \\
Frequency resolution & 24.4 kHz \\
Flux calibrators & 3C48 and 3C286 \\
                 & ($01^h 37^m 41^s + 33^\circ 09' 35''$) \\ 
                 & ($13^h 31^m 08^s + 30^\circ 30' 33''$)\\ 
Phase calibrator & $2321-163$ \\
                 & ($23^h 21^m 02^s - 16^\circ 23' 05''$) \\ 
Pointing center  & GAMA-23 \\
                 & ($23^h 00^m 00^s - 32^\circ 29' 60''$) \\ 
\hline
\end{tabular}
\end{table}
The uGMRT is a radio interferometer made of dish antennas situated in Khodad, Pune, India. This observation was organised after the upgrade of the wideband system in uGMRT. We observed the GAMA-23 field ($\alpha_\text{J2000} = 23\text{h}00\text{m}00\text{s}$, $\delta_\text{J2000} = -32.50^{\circ}$) for Band-2 ($120-250$\,MHz). The observation was conducted in the observing cycle 42 with a total observation time of 9 hours. The details of observation is mentioned in the Table~\ref{table:tab1}. The GAMA-23 field is located at galactic latitudes ($\ell = 13.61\deg$, $b = -65.31\deg$), which is away from the Galactic plane. The wideband data in Band-2 ($120-250$\,MHz) is restricted by a notch filter\footnote{\url{http://www.ncra.tifr.res.in:8081/~secr-ops/etc/etc_help.pdf}} for frequency range $165-190$\,MHz, so the usable bandwidth for this band is $105$\,MHz without processing any calibration. However, Radio Frequency Interference (RFI) is strong at lower frequencies overall, so we opt for real-time broadband RFI filtering with a 4$\sigma$ cut-off for this deep-field observation. \citep{2017arXiv171008576B, 2022JAI....1150008B}. We observed two flux calibrator sources 3C286 and 3C48 for calibration of the target according to their availability through the observation. We also observe the phase calibrator (2321-163) to account for phase referencing.

\subsection{Data Reduction}\label{sec:data_reduction}

The data calibration has been performed using Source Peeling and Atmospheric Modelling \citep[{\tt\string SPAM};][]{2009A&A...501.1185I,  2014ASInC..13..469I}. SPAM is a Python script based on the Astronomical Image Processing System \citep[{\tt\string AIPS};][]{2003ASSL..285..109G}. This is a semi-automated pipeline that accounts for ionospheric phase delay. \\
\begin{figure*}
\includegraphics[width=0.75\paperwidth]{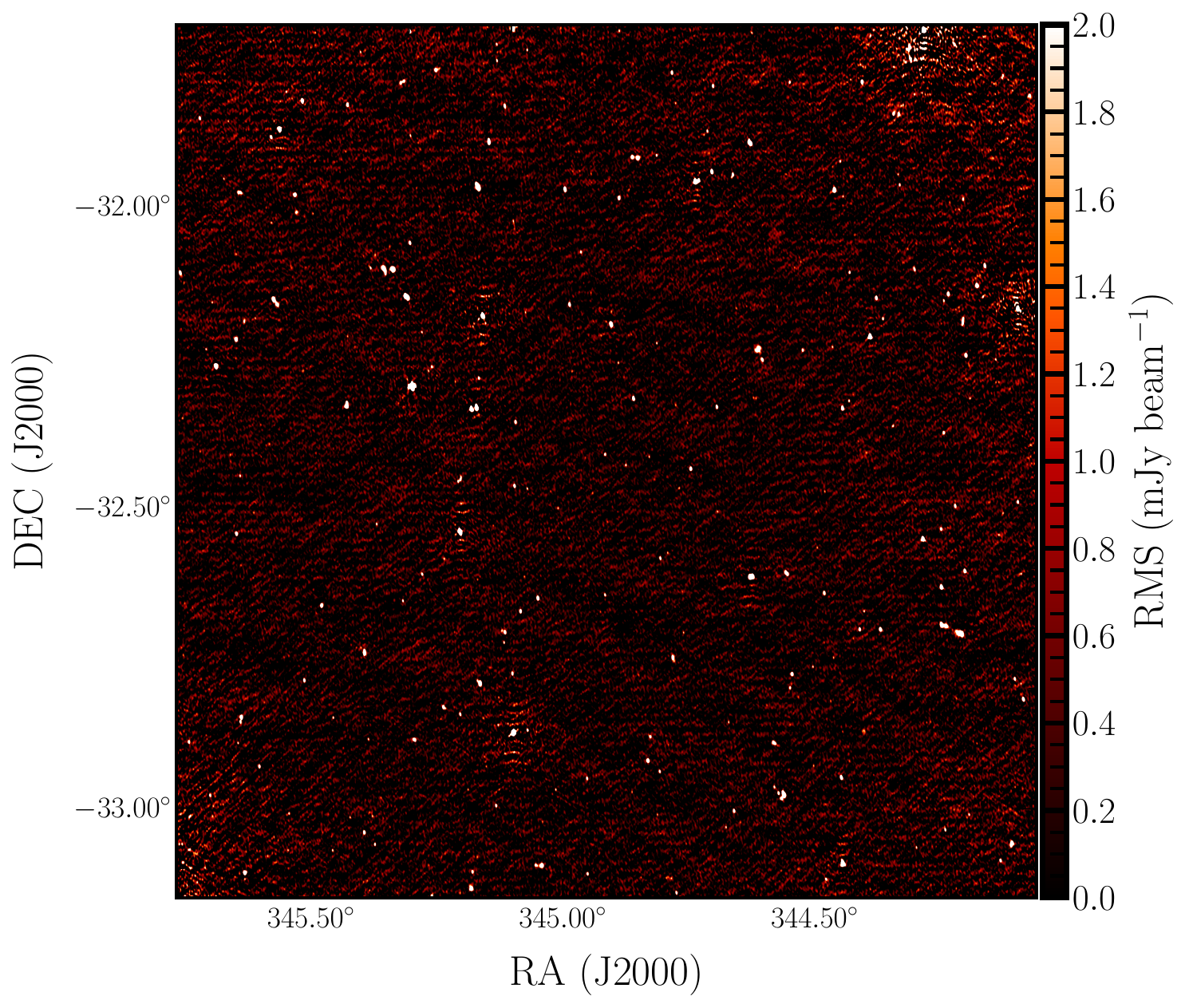}
\caption{The central region of the final image of GAMA-23 field with uGMRT at 183\,MHz. The area of the shown region is $1.41\,\text{deg}^2$. The off-source RMS noise and restoring beam of the final image are 320\,$\mu$Jy beam$^{-1}$ and $15.4'' {\times} 8.2''$, respectively.}
\label{fig:Fig1}
\end{figure*}
\begin{figure*}
\centering\includegraphics[width=0.48\linewidth]{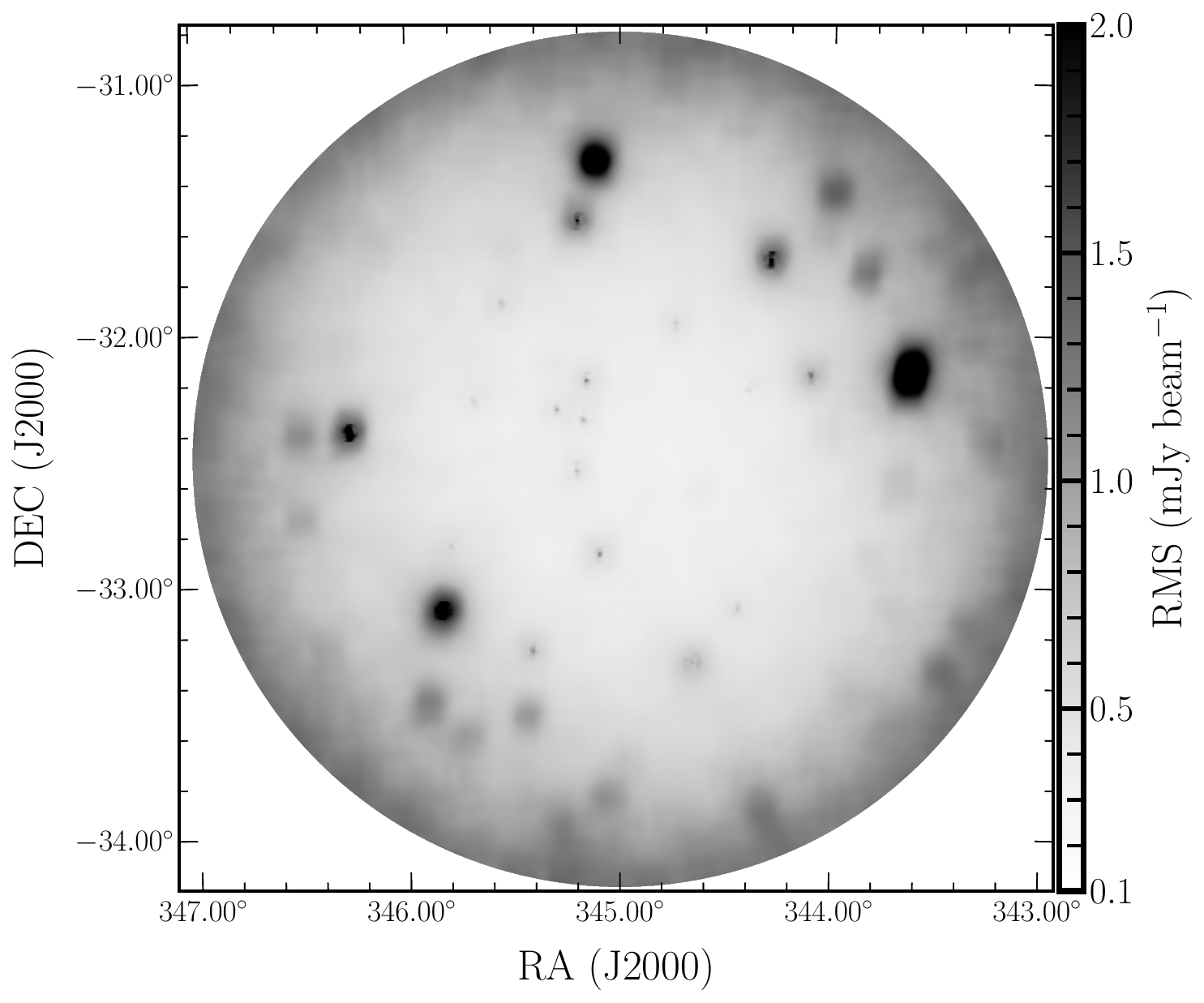}
\hfill
\centering\includegraphics[width=0.45\linewidth]{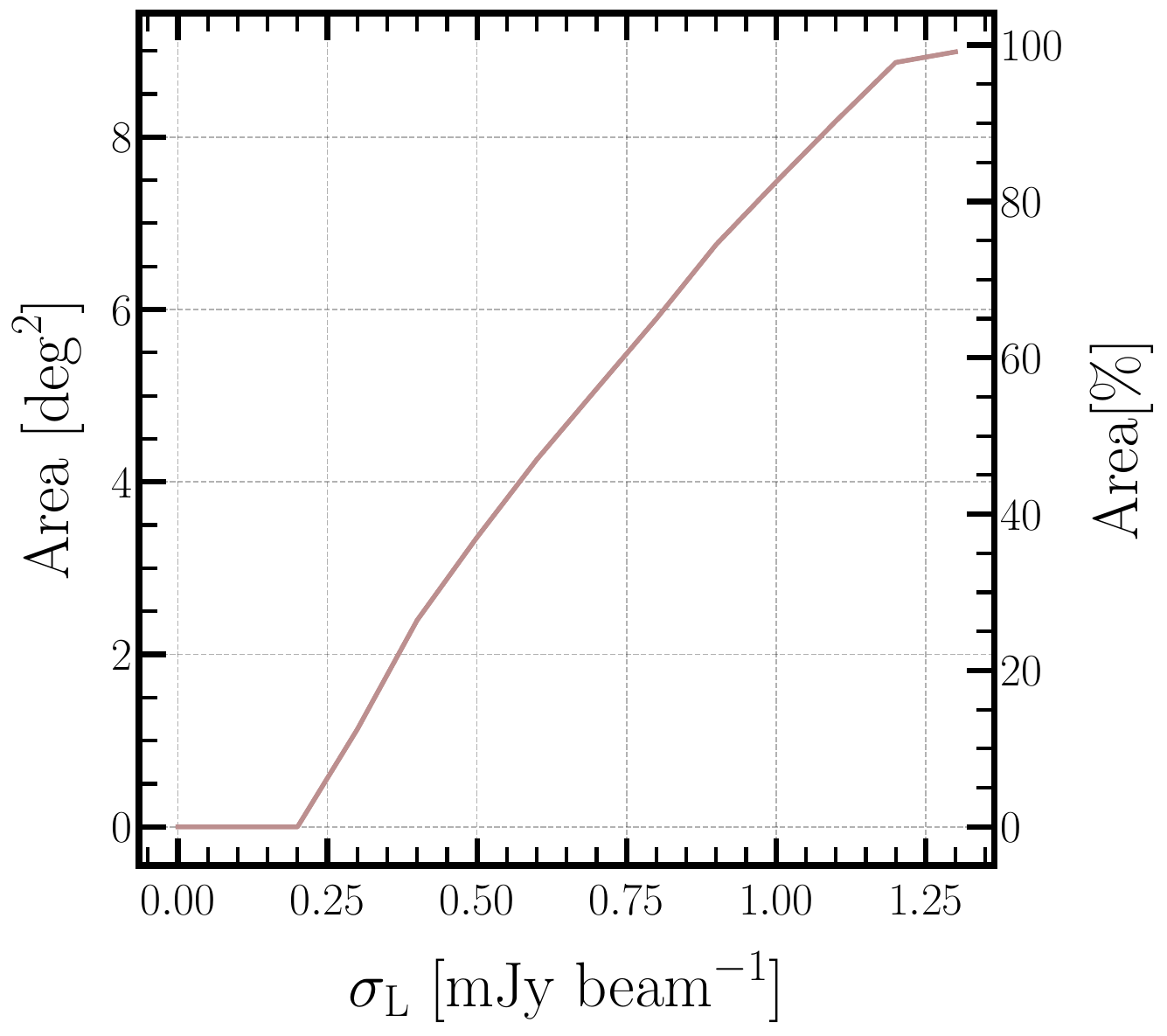}
\caption{The background RMS noise map of the final image of the GAMA-23 field (\textit{left}). Cumulative area of final image with respect to local RMS noise (\textit{right}).}
\label{fig:Fig2}
\end{figure*}
SPAM divides the wideband data into small subbands of 16\,MHz bandwidth to process them easily. In the precalibration stage, the SPAM flags bad antennas, RFI and corrupt data. The flux scale model \cite{2012MNRAS.423L..30S} was used to set the flux density of the primary calibrator. The gain and bandpass solutions have been calculated in this procedure, and each subband is averaged over time and frequency to reduce volume before processing further \citep[see][]{2025MNRAS.544.3617S}. \\ 
These subbands are processed independently in the main SPAM pipeline. This procedure includes the iterative wide-field imaging and self-calibration. In this step, we use the catalogue (reference sky model) generated from calibrated narrowband data, which was observed simultaneously with wideband data. Following the direction-independent calibration, the ionospheric phase corrections were applied using the peeling method around the bright sources. The final calibrated visibilities from each subband were combined to produce the final image.

\subsection{Imaging}\label{sec:imaging}

We used {\tt\string WSClean}\footnote{\url{https://gitlab.com/aroffringa/wsclean}} \citep{2014MNRAS.444..606O} to produce the final continuum image of wideband data. We used multiscale cleaning for deconvolution \citep{2017MNRAS.471..301O} and the Briggs parameter \citep{1995PhDT.......238B} for uniform weighting during imaging. We corrected for primary beam using {\tt\string uGMRTprimarybeam}\footnote{\url{https://github.com/ruta-k/uGMRTprimarybeam}} for uGMRT. We used 20\% threshold to correct for the primary beam effect. The final primary beam corrected image has off-source RMS noise 320\,$\mu$Jy beam$^{-1}$. Here, the off-source RMS noise is the RMS noise measured in the source free region of the central area of the image. The dynamic range of the image $S_\text{peak}/{\sigma} = 1260$. The restoring beam size is  $15.4'' \times 8.2''$, and the position angle is $6.7^\circ$. The central region of the final image is shown in Figure~\ref{fig:Fig1}. The final image covers an area of $5.86\,\text{deg}^2$ while the central region shown covers an area of $1.41\,\text{deg}^2$.

\section{Source Catalog} \label{sec:catalogue}

We used a source extractor {\tt\string PyBDSF}\footnote{\url{https://www.astron.nl/citt/pybdsf/}} \citep{2015ascl.soft02007M} to produce a catalogue from the final image obtained from section~\ref{sec:imaging}. We used {\tt\string rms\_box} of size ($180, 50$). We used adaptive setting and selected a small rms box of size ($35, 7$) near the bright sources to reduce the false detection rate. We used $150{\sigma}_\text{RMS}$ adaptive threshold to select these bright regions. We selected {\tt\string psf\_vary\_do = True} to account for varying PSF across the image. We used {\tt\string thres\_isl $= 3\sigma$} and {\tt\string thres\_pix $= 5\sigma$} as the island and source detection thresholds, respectively. The produced catalogue consists of 569 sources above 1.6\,mJy ($5{\sigma}$). The sample of catalogue sources is presented in Table~\ref{tab:gama23_catalog} and the electronic version will be available online with the paper.

\begin{table*}
\centering
\caption{Sample sources from the final source catalogue of the GAMA-23 field image using uGMRT at Band-2 (120-250\,MHz). The produced source catalogue consists of the columns listed below, corresponding to source IDs, source positions (RA and DEC), total and peak flux densities, source sizes - major (Maj) and minor (Min) axes, positional angle (PA), and the RMS noise of the fitted islands.}
\label{tab:gama23_catalog}
\resizebox{\textwidth}{!}{
\begin{tabular}{ccccccccc}
\hline
Source ID & RA & DEC & Total\_Flux & Peak\_Flux & Maj & Min & PA & Isl\_rms \\
& (deg $\pm$ arcsec) & (deg $\pm$ arcsec) & (mJy) & (mJy beam$^{-1}$) & (arcsec) & (arcsec) &(deg) & (mJy beam$^{-1}$) \\
\hline
0 & 346.9942 $\pm$ 1.19 & -32.3103 $\pm$ 0.95 & 17.84 & 9.00 & 17.07 & 14.66 &  102.31 & 1.31 \\
1 & 346.8557 $\pm$ 1.01 & -30.8626 $\pm$ 1.28 & 10.39 & 7.30 & 16.68 & 10.76 &  34.05 & 1.24 \\
2 & 346.7789 $\pm$ 0.40 & -32.1506 $\pm$ 0.46 & 24.05 & 16.42 & 15.24 & 12.13 &  33.12 & 1.11\\ 
\hline
\end{tabular}
}
\tablenotes{The full catalogue will be available in the form of a table in ``.FITS'' format with the electronic version of this paper.}
\end{table*}

We produced the background RMS noise map of the final image. In the left panel of Figure~\ref{fig:Fig2}, the RMS noise map highlights the affected region near the bright source and residual artefacts. The right panel of the Figure~\ref{fig:Fig2} represents the variation of noise with respect to the image area and the corresponding percentage of the field. The noise increases as we move from the centre to the edge of the image due to a decline in the primary beam sensitivity. 

The confusion noise limit also limits the accuracy of the source catalogue. This noise emerges due to multiple faint, unresolved sources within the beam. At low frequency, as the beam is large and the resolution is low, confusion noise becomes important. However, in our case, the observation time is limited, the RMS noise is larger than the confusion noise (320\,$\mu$Jy beam$^{-1}>$ 76\,$\mu$Jy beam$^{-1}$) \citep[equation 1;][]{2025MNRAS.544.3617S}.

\subsection{Source Classification} \label{sec:classification}

\begin{figure}
\includegraphics[width=3.35in]{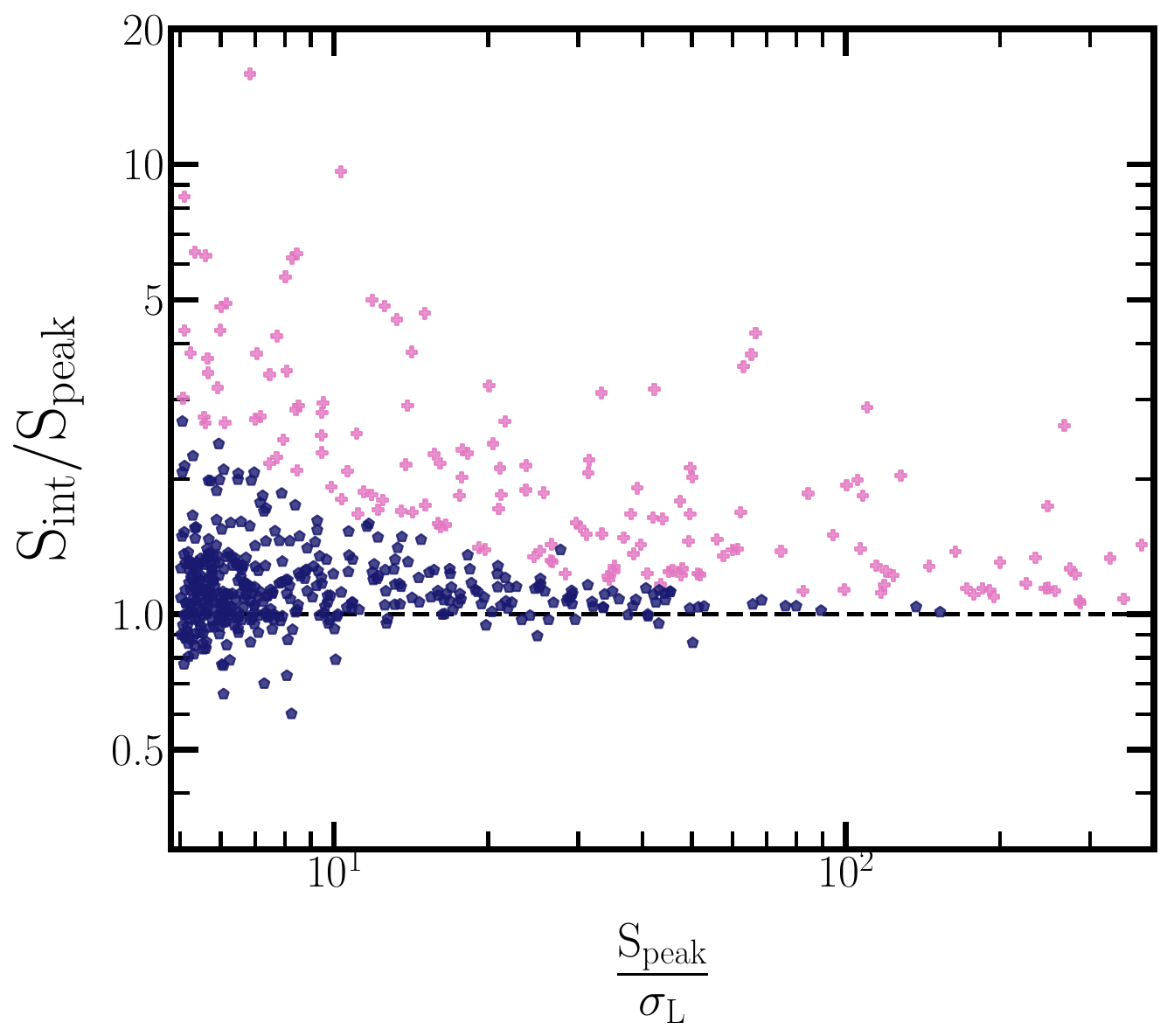}
\caption{Integrated to peak flux density ratio ($S_{\text{int}}/S_{\text{peak}}$) with respect to signal-to-noise ratio ($S_{\text{peak}}/\sigma_L$) for uGMRT catalogue sources. Point sources are shown as blue pentagons, and resolved sources are presented with pink plus signs.}
\label{fig:Fig3}
\end{figure}

Various effects, such as time and bandwidth smearing, may elongate the source in the image plane, which makes it difficult to classify resolved and point like sources. However, the integrated flux densities are not necessarily affected by time smearing, although it can reduce the peak brightness. Thus, using the ratio ${S_{\text{int}}}/{S_{\text{peak}}}>1$ alone is not sufficient for the classification of resolved and point sources. We present Figure~\ref{fig:Fig3} to show the variation of the ratio ${S_{\text{int}}}/{S_{\text{peak}}}$ with respect to ${S_{\text{peak}}}/{{\sigma}_{\text{L}}}$. The figure shows skewed distribution at low SNR, which could be due to calibration uncertainties and local noise variation. We followed \cite{2015MNRAS.453.4020F, 2019PASA...36....4F} to classify the resolved and point like sources in our catalogue, as described below:
\begin{equation}
\label{eqn:2}
\sigma_\text{RMS} = \sqrt{\left(\frac{\sigma_{S_{\text{int}}}}{S_{\text{int}}}\right)^2 + \left(\frac{\sigma_{S_{\text{peak}}}}{S_{\text{peak}}}\right)^2}
\end{equation}
Based on the described criteria, a source will be resolved, for $\text{ln}({S_{\text{int}}}/{S_{\text{peak}}})> 3{\sigma}_\text{RMS}$ \citep{2015MNRAS.453.4020F}. Thus, we have found 159 resolved sources (pink plus signs) and 410 point like sources (blue pentagons), as presented in Figure \ref{fig:Fig3}. 
\begin{figure}[t]
\includegraphics[width=3.35in]{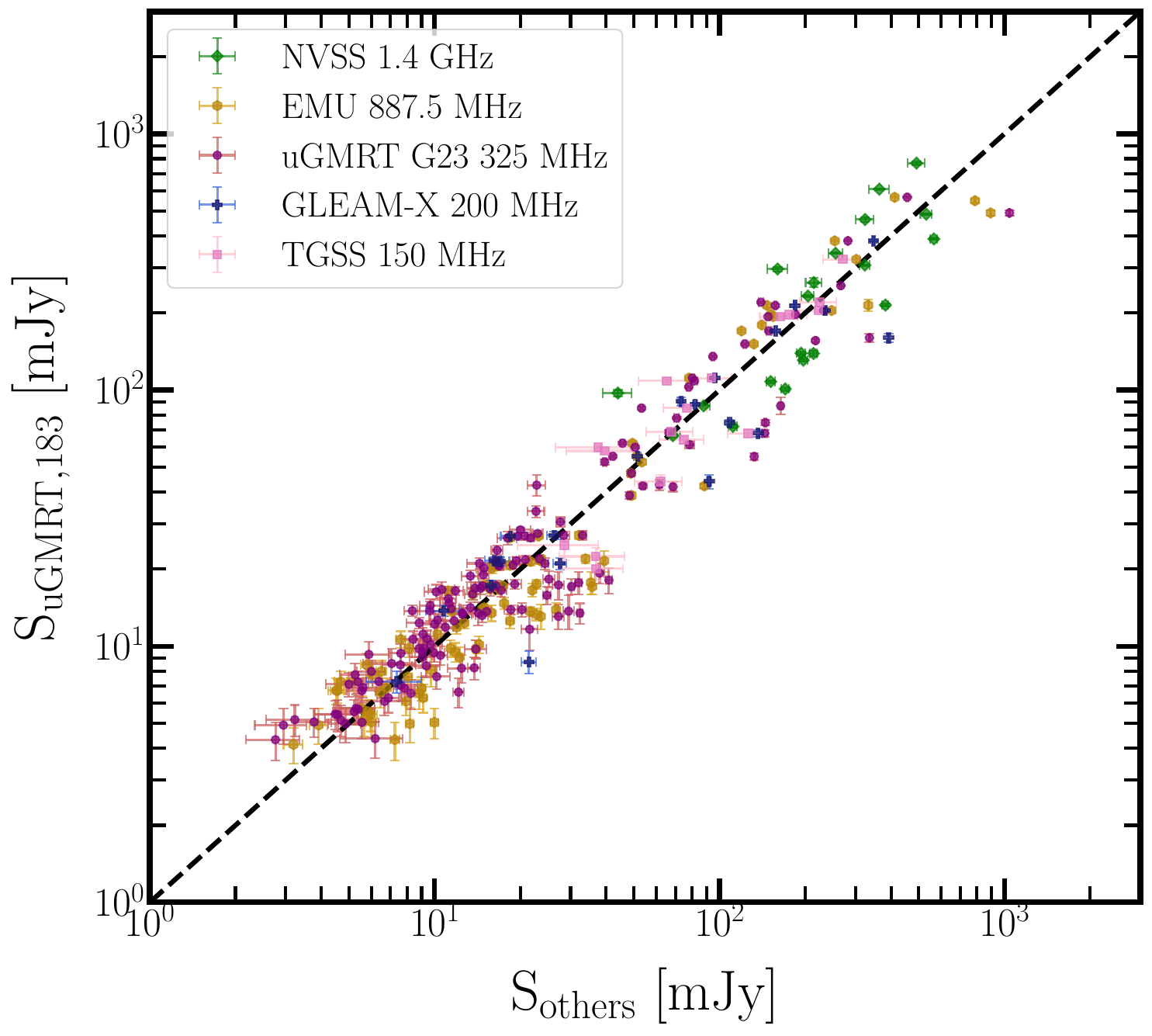}
\caption{Comparison of the integrated flux densities of sources computed from the uGMRT catalogue at 183\,MHz with other catalogues after rescaled at 183\,MHz. The matched sources follow the criteria of high SNR, compact and isolated sources. The black dashed line shows the $S_{\text{uGMRT}}/S_{\text{others}} = 1$ while the ratios of sources with other catalogues presented are NVSS 1400\,MHz (green), ASKAP EMU 887.5\,MHz (golden), uGMRT 325\,MHz (purple), GLEAM-X 200\,MHz (blue) and GMRT TGSS 150\,MHz (pink).}
\label{fig:Fig4}
\end{figure}

\subsection{Comparison with other radio catalogues} \label{sec:comparison}

We present a comparison of our 183\,MHz catalogue with previously available catalogues of the GAMA-23 field. The comparison is important to verify the reliability of our catalogue. Uncertainties in the uGMRT primary beam model can introduce uncertainties in the measured flux densities of catalogue sources. non-ideal ionospheric corrections can leave residual phase errors, leading to a positional offset in the detected sources. We used GMRT TGSS 150\,MHz \citep{2017A&A...598A..78I}, GLEAM-X 200\,MHz \citep{2022PASA...39...35H}, uGMRT 325\,MHz \citep{2026AJ....172..197S}, ASKAP EMU survey 887.5\,MHz \citep{2022MNRAS.512.6104G} and NVSS 1400\,MHz \citep{1998AJ....115.1693C} for comparison. 
\begin{figure}
\includegraphics[width=3.35in]{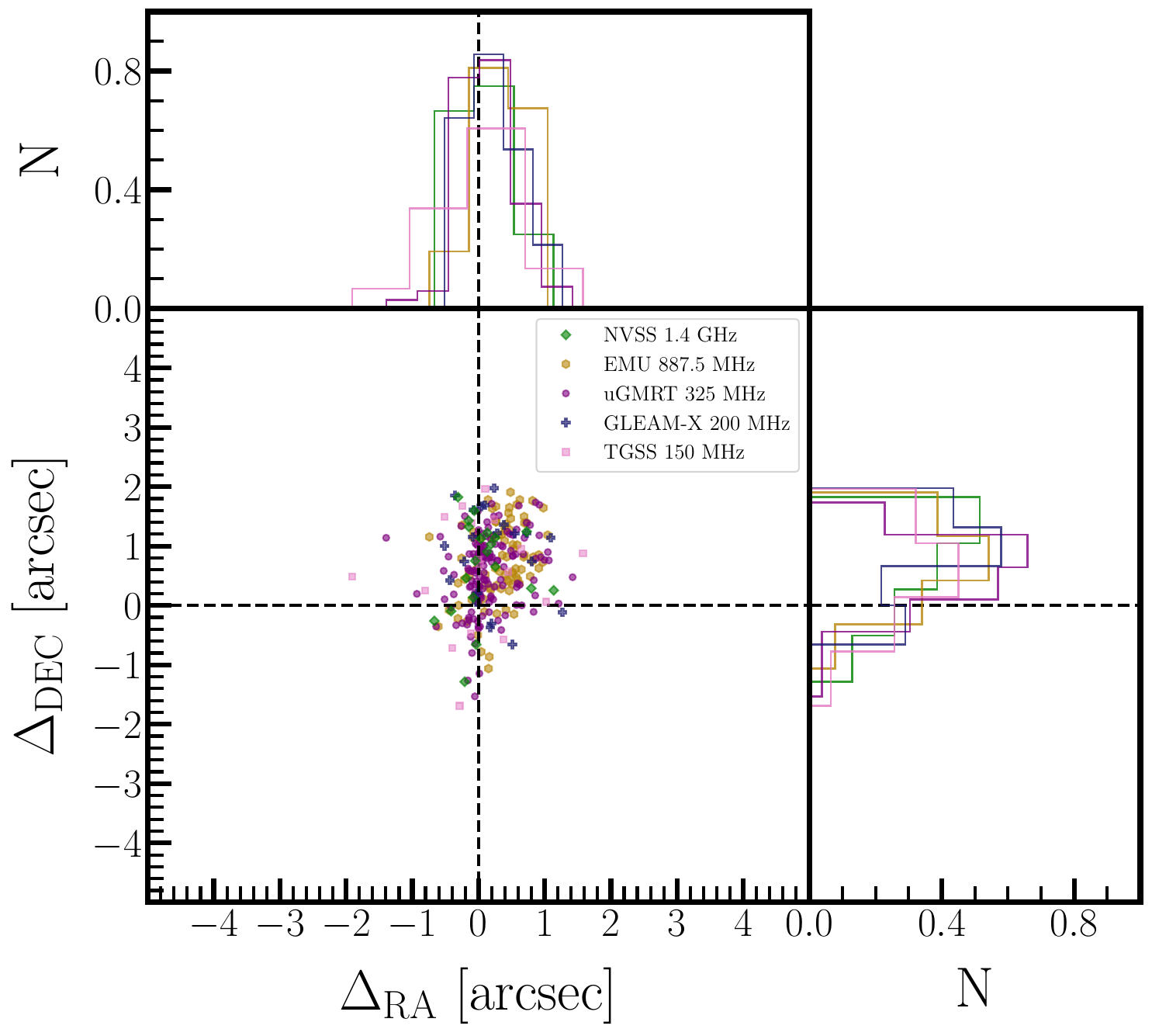}
\caption{Positional offsets in RA $\&$ DEC of the uGMRT sources at 183\,MHz compared with NVSS 1400\,MHz (green), ASKAP EMU 887.5\,MHz (golden), uGMRT 325\,MHz (purple), GLEAM-X 200\,MHz (blue) and GMRT TGSS 150\,MHz (pink) catalogues.}
\label{fig:Fig5}
\end{figure}
\begin{figure}
\includegraphics[width=3.35in]{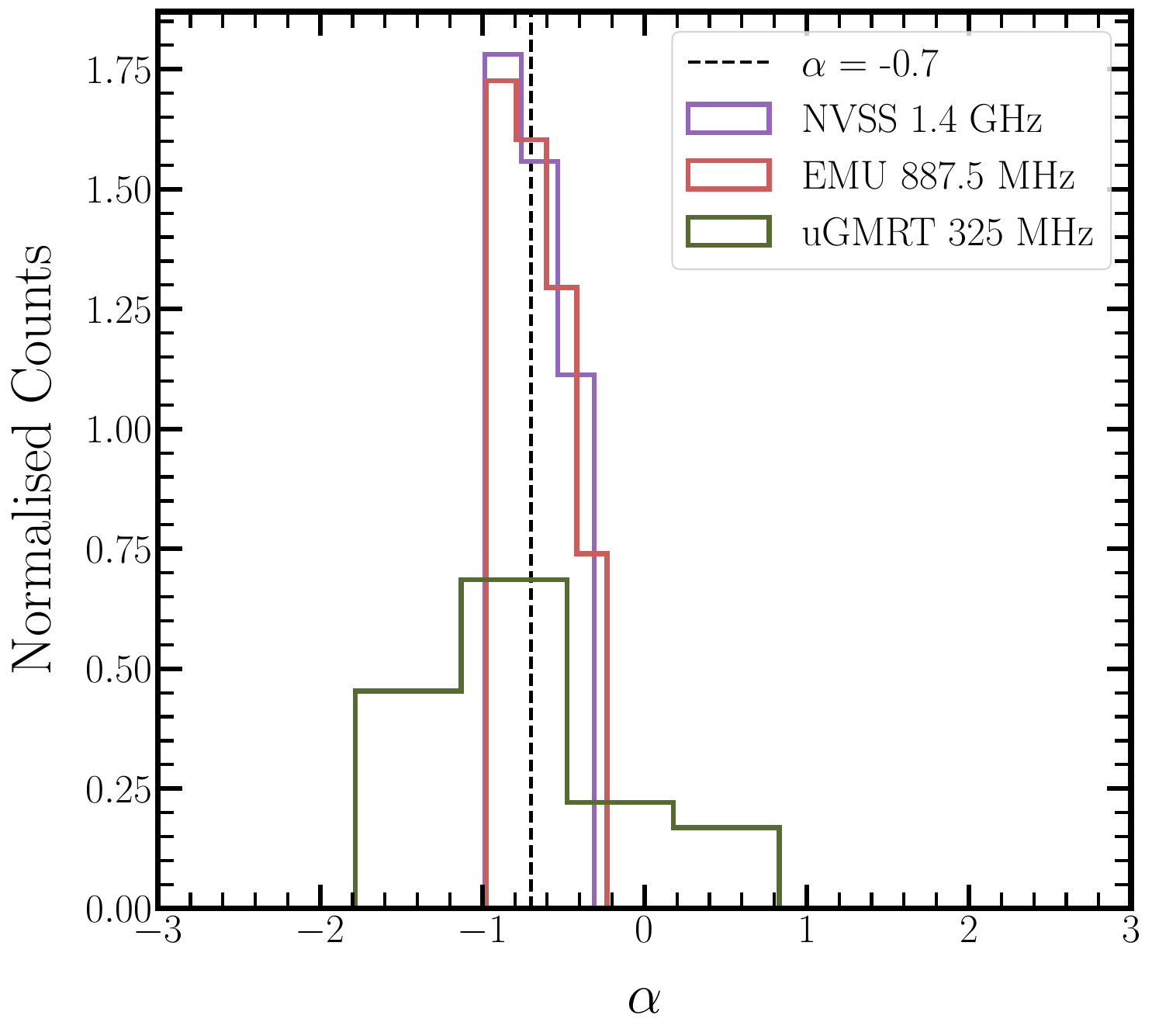}
\caption{The spectral indices estimated in the GAMA-23 field in comparison with 1400\,MHz NVSS (purple), 887.5\,MHz ASKAP EMU (indianred) and 325\,MHz uGMRT (olivegreen) as histograms after using a matching radius of $25''$, $15''$ and $20''$ for NVSS, ASKAP EMU and uGMRT respectively. The assumed value of the spectral index for the power law {${\alpha} = -0.7$} is shown with the black dashed line.}
\label{fig:Fig6}
 \end{figure}
Each catalogue has their own flux limit (S$_\text{limit}^{\dagger}$) based on their completeness and sensitivity. We have used the flux densities of sources above S$_\text{limit}^{\dagger}$ for each catalogue as mentioned in Table~\ref{table:fluxlimit}. The search radius to match the counterpart in the other catalogue is set to $1.5-2.0\,{\times}$\,PSF, 
as shown in Table~\ref{table:fluxlimit}.

\begin{table*}
\centering
\caption{The catalogues for comparison correspond to their frequency, resolution, and flux limit. The flux limit after scaling at 183\,MHz (assumption $\alpha = -0.7$). The median positional offsets in RA and declination for the uGMRT 183\,MHz catalogue relative to previous catalogues and their respective errors at 16th and 84th percentile. Finally, the number of matched sources of our catalogue after comparison with other catalogues.}
\resizebox{\textwidth}{!}{
\begin{tabular}{ccccccccc}
\hline
Catalogue & Frequency & Resolution & $S_\text{limit}^{\dagger}$ & $S_\text{{cut}}^{\text{183}}$ & Cross-match radius & $\delta_\text{RA,median}$ & $\delta_\text{DEC,median}$ & Match \\ [0.5ex]
 & (MHz) & (arcsec) & (mJy) & (mJy) & (arcsec) & (arcsec) & (arcsec)  & \\ [1ex]
\hline
uGMRT & 183 & 15.4 & 1.6 & 1.6 & - & - & - & - \\ [0.5ex]
NVSS & 1400 & 45 & 2.0 &  8.31 & 25 & -0.041$^{+0.17}_{-0.29}$ & 0.830$^{+0.92}_{-0.49}$ & 20\\ [0.5ex]
TGSS & 150 & 25 & 17.5 &  15.22 & 25 & -0.005$^{+0.44}_{-0.54}$ & 0.698$^{+1.21}_{-0.77}$ & 17 \\ [0.5ex]
uGMRT & 325 & 15.8 & 0.55 & 0.82 & 20 & 0.077$^{+0.19}_{-0.50}$ & 0.590$^{+0.70}_{-0.57}$ & 145 \\ [0.5ex]
EMU &   887.5 & 10 & 0.2 &  0.6 & 15 & 0.361$^{+0.44}_{-0.30}$ & 0.628$^{+0.54}_{-0.77}$ & 87 \\ [0.5ex]
GLEAM-X & 200 & 45 & 3.5 &  3.72 & 70 & 0.179$^{+0.37}_{-0.51}$ & 1.143$^{+1.22}_{-0.52}$ & 21  \\ [0.5ex]
\hline
\end{tabular}}
\label{table:fluxlimit}
\tablenotes{$S_\text{limit}^{\dagger}$ - Flux density limit for selected catalogues for comparison.}
\end{table*}

\subsection{Flux Density Offset} \label{sec:fluxoffset} 

We have compared the flux densities of sources in the uGMRT catalogue for the GAMA-23 field at 183\,MHz with those in previous available catalogues. We adopted the flux scale from \cite{2012MNRAS.423L..30S} and scaled the flux densities of other catalogues to this flux scale. We have used the NVSS catalogue at 1400\,MHz, the ASKAP EMU catalogue at 887.5\,MHz, the uGMRT catalogue at 325\,MHz, the GLEAM-X catalogue at 200\,MHz and the TGSS catalogue at 150\,MHz. \\
The selection criteria for matching the source from the counterpart were that sources should have a high signal-to-noise ratio ($>10{\sigma}$), be compact, and isolate sources \citep[see;][]{2023MNRAS.525.5311S}. We scaled other catalogues at different observing frequencies to 183\,MHz using $S_\nu \propto \nu^{\alpha}$ with the spectral index $\alpha=-0.7$.
The Figure~\ref{fig:Fig4} shows the variation of the flux density ratios of the matched sources of this catalogue in comparison with other previous catalogues. $S_{183 \text{MHz}}$ presents our catalogue while S$_{\text{others}}$ stands for other catalogues used for comparison. The black dashed line represents the $S_{183 \text{MHz}}/S_{\text{others}}=1$. The median ratios obtained are 1.01$^{+0.30}_{-0.34}$ (TGSS 150\,MHz), 1.07$^{+0.54}_{-0.19}$ (GLEAM-X 200\,MHz), 1.10$^{+0.39}_{-0.29}$ (uGMRT 325\,MHz), 0.97$^{+0.29}_{-0.33}$ (EMU 887.5\,MHz) and 1.04$^{+0.40}_{-0.50}$ (NVSS 1400\,MHz) with 16th and 84th percentile errors. 

\subsection{Positional Offset} \label{sec:pos-off} 

The positional offset of the catalogue sources is obtained by comparing the position of the matched sources. For comparison, we have used the NVSS 1400\,MHz, ASKAP 887.5\,MHz, uGMRT 325\,MHz, GLEAM-X 200\,MHz, and TGSS 150\,MHz catalogues. We have adopted similar criteria to match sources, as mentioned in Section~\ref {sec:fluxoffset}. Positional offset has been calculated using the formula from \cite{2016MNRAS.460.2385W}:
\begin{equation}
\label{eqn:3}
\begin{aligned}
\delta_{\rm RA}  &= {\rm RA}_{\rm uGMRT_{183}} - {\rm RA}_{\rm NVSS} \\
\delta_{\rm DEC} &= {\rm DEC}_{\rm uGMRT_{183}} - {\rm DEC}_{\rm NVSS}
\end{aligned}
\end{equation}
The Figure~\ref{fig:Fig5} represents the positional offset of matched sources in the form of histograms in comparison with other catalogues. Table~\ref{table:fluxlimit} presents the deviation in RA and Dec of our catalogue with respect to other catalogues. The deviation is within the limit of the cell size of the image, i.e., $2.5''$. The highest-frequency catalogue was NVSS at 1400\,MHz to correct for ionospheric fluctuations. However, the NVSS catalogue has a resolution of $45''$ while our catalogue has a resolution of $15.4''$. Thus, we have not applied any positional correction in the final catalogue.

\subsection{Spectral Index Distribution}

The GAMA-23 deep field is a source-rich field. We have characterised the spectral properties of the sources using the flux densities extracted from our uGMRT source catalogue. The spectral index ${\alpha}$ has been computed from the matched sources after comparison with the NVSS 1400\,MHz catalogue and the ASKAP EMU at 887.5\,MHz catalogue using the power law distribution $S_\nu \propto \nu^{\alpha}$. We have calculated the spectral index median and presented its distribution in the form of histograms, shown in Figure~\ref{fig:Fig6}. The calculated median spectral indicies, ${\alpha}_\text{median}$ are $-0.72^{+0.19}_{-0.24}$ (NVSS 1400\,MHz), $-0.68^{+0.18}_{-0.23}$ (ASKAP EMU 887.5\,MHz) and $-0.86^{+0.40}_{-0.77}$ (uGMRT GAMA-23 325\,MHz) with their respective errors for the 16th and 84th percentile. Thus, the ${\alpha}_\text{median}$ is approx ${-0.7}$ with the uncertainty for our catalogue. The spectral indices of our source catalogue derived using NVSS ($\alpha_{1400}^{183}$) are consistent with the results of \cite{2016MNRAS.463.2997M}, which reported a median spectral index of $\alpha_{1400}^{150}=-0.78\pm0.015$ down to a flux density of 1.9\,mJy. This suggests that the source population is dominated by star-forming galaxies with non-thermal radio emission \citep{2024MNRAS.528.5346A}. However, a detailed study of the classification between AGNs and SFGs, including the steep spectrum sources, will be presented in future work.

\section{Source Count} \label{sec:count}

Radio source counts provide a statistical analysis of the extragalactic radio source population \citep{1969MNRAS.142....1R, 2001A&A...365..392P, 2023MNRAS.523.5187W}. It is widely used to study galaxy evolution, the growth of supermassive black holes and their relative roles in star formation and AGN activity \citep{best2005host, 2010PNAS..107.7184B, 2017NatAs...1E.165H}. Different source populations dominate at different flux densities, source counts can be combined with multi-wavelength data to trace how AGN incidence, star formation, and the surrounding environment affect evolution with cosmic time \citep{2023MNRAS.523.5187W, 2024FPP....1100059F}. Source counts provide information about the evolutionary model of the source population by comparing observations with predicted models \citep{1998MNRAS.296..839H, 2020MNRAS.491.1127O, 2021A&A...648A...5M}. \\
We have estimated the differential source count for the GAMA-23 field for the first time at 183\,MHz down to 1.7\,mJy. However, the source extractor PyBDSF has its own limitations, which can lead to incompleteness and biases in the raw source counts. We have discussed and corrected for the two major biases, one due to false detection rate and the other due to incompleteness in the catalogue below, following the criteria described in \citep{2016MNRAS.460.2385W}.  

\subsection{False Detection Rate}

\begin{figure}
\includegraphics[width=3.35in]{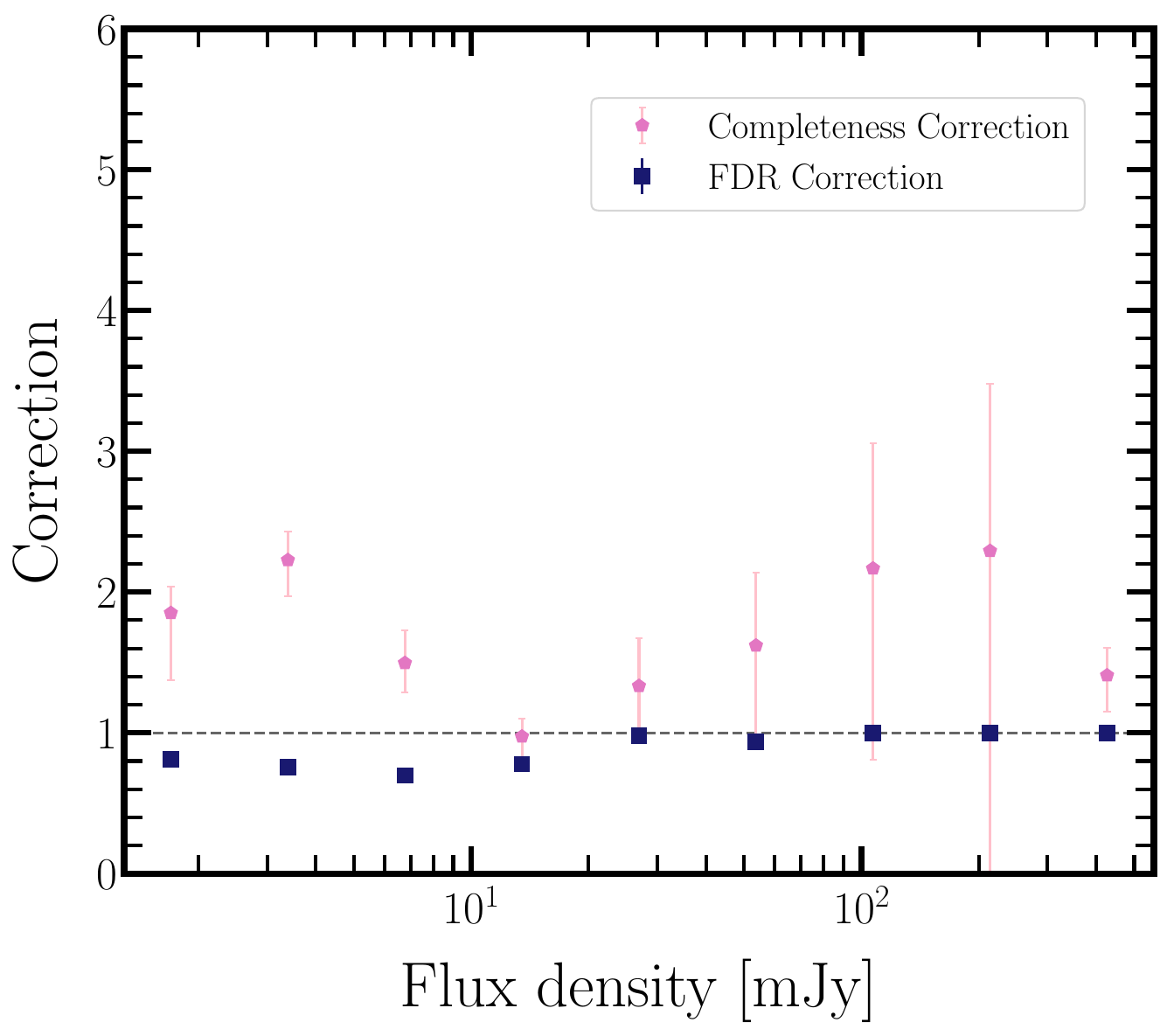}
\caption{Median values and their corresponding errors for the correction factors of false detection rate (FDR) (navy) and completeness (pink) with integrated flux density for each flux bin.}
\label{fig:Fig7}
\end{figure}

The possibility of false detection rate (FDR) arises due to artefacts and noise in the extracted catalogue from PyBDSF. The number of spurious sources that are detected as real is known as the false detection rate. Assuming symmetric noise around zero, we have estimated the catalogue of negative-noise-peak sources from the inverted image using PyBDSF with the same parameters as for the original image. We have detected 116 sources with negative peaks ${\leq}-5{\sigma}$ threshold. Following the method described in the \cite{2019A&A...622A...4H}, we have estimated the fraction of real sources in each flux density bin using Equation 4 provided in \cite{2026AJ....172..197S} and multiplied the fraction together with completeness (discussed in section~\ref{sec:completeness}) by the raw source counts to obtain the corrected source counts. Figure~\ref{fig:Fig7} represents the sources extracted from the inverted image in the form of a catalogue, binned into the 9 logarithmic flux density bins and represented by navy squares.

\begin{figure*} 
\makebox[\textwidth]{\includegraphics[width=0.75\paperwidth]{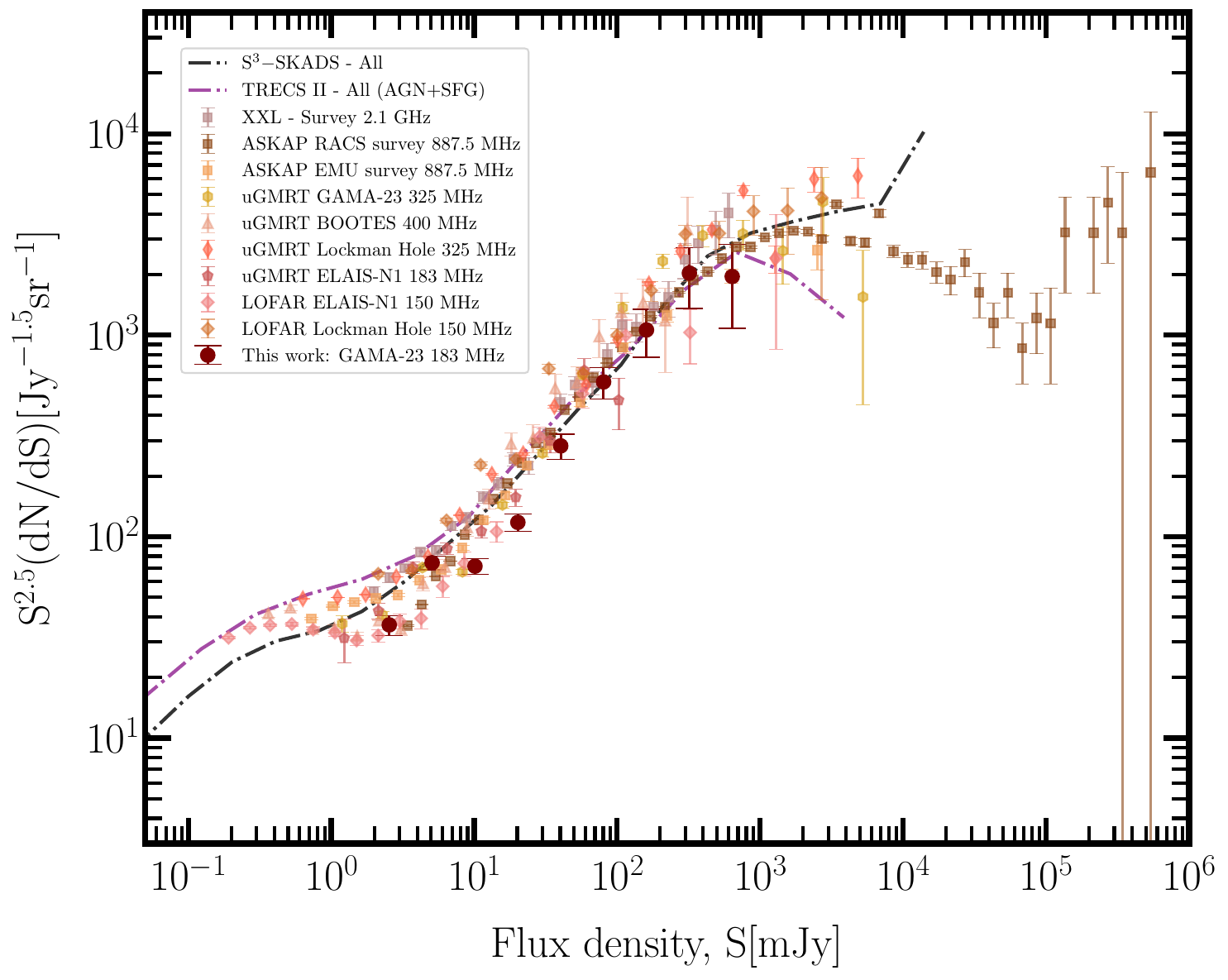}}
\caption{The Euclidean-normalised differential corrected source counts at 183\,MHz for the GAMA-23 field are presented in maroon circles. We have multiplied by FDR and the incompleteness correction factors for each flux bin. We used simulation predicted models, the S$^3$-SKADS \citep{2008MNRAS.388.1335W} and T-RECS II \citep{2023MNRAS.524..993B}, for comparison, represented by black and purple dot-dashed lines, respectively. The other previous observations and surveys are ATCA XXL$-$South field at 2.1\,GHz \citep{2018A&A...620A...3B} (rosybrown squares), ASKAP RACS survey at 887.5\,MHz \citep{2021PASA...38...58H} (saddlebrown squares), ASKAP-EMU (GAMA-23) at 887.5\,MHz \citep{2022MNRAS.512.6104G} (sandybrown squares) and uGMRT GAMA-23 field at 325\,MHz \citep{2026AJ....172..197S} (golden hexagons), covering the southern sky. The other deep fields considered for comparison are Bo\"{o}tes 400\,MHz uGMRT \citep{2023MNRAS.525.5311S} (salmon triangles), Lockman Hole 325\,MHz GMRT \citep{2020MNRAS.495.4071M} (orange diamonds), ELAIS-N1 - 183\,MHz uGMRT \citep{2025MNRAS.544.3617S} (indianred pentagons), LoTSS (ELAIS-N1) 150\,MHz LOFAR \citep{2021A&A...648A...5M} (lightcoral diamonds) and Lockman Hole 150\,MHz LOFAR \citep{2016MNRAS.463.2997M} (chocolate diamonds).
}
\label{fig:Fig8}
\end{figure*}

\subsection{Completeness} \label{sec:completeness}

The completeness of the catalogue can be affected by two or more biases, which can lead to overestimation and underestimation of the sources in the extracted catalogue. Catalogue incompleteness occurs when the source extractor fails to identify real sources due to sensitivity limits, noise variation across the image, extended sources that cause differences in total-to-peak flux, or imaging artefacts near bright sources. Around the detection threshold, Eddington bias \citep{1913MNRAS..73..359E} causes an excess of flux in faint source count bins by redistributing faint sources into higher flux density bins, while resolution bias \citep{2006A&A...457..517P} occurs when extended sources whose peak flux density falls short of the detection threshold. Both effects are corrected using injection recovery simulations that support flux deboosting and completeness modelling.\\

\begin{table*}
\centering
\caption{Derived Euclidean normalised differential source counts of the GAMA-23 field at 183\,MHz. The given columns correspond to the flux density range and its central value for each bin, normalised raw source counts (S$^{2.5}$(dN/dS)), median value corresponding to each bin for FDR and completeness correction factors, and normalised corrected source counts with their associated Poisson errors.}
\label{table:dNdS}
\resizebox{\textwidth}{!}{
\begin{tabular}{ccccccc}
\hline
$S_{\rm range}$ & $S_{\rm mid}$ & N & $S^{2.5}(\text{dN/dS})$ & FDR & Completeness & Corrected $S^{2.5}(\text{dN/dS})$ \\
(mJy) & (mJy) & & (Jy$^{1.5}$ sr$^{-1}$) & & & (Jy$^{1.5}$ sr$^{-1}$) \\
\hline
1.700\,--\,3.391 & 2.546 & 80 & 24.343 $\pm$ 2.722 & 0.81 $\pm$ 0.19 &  1.85$^{+0.19}_{-0.48}$  & 36.498 $\pm$ 4.081 \\
3.391\,--\,6.764 & 5.078 & 160 & 43.892 $\pm$ 3.470 & 0.76 $\pm$ 0.24 & 2.23$^{+0.20}_{-0.26}$ & 46.401 $\pm$ 3.668 \\
6.764\,--\,13.493 & 10.129 & 122 & 68.188 $\pm$ 6.174 & 0.70 $\pm$ 0.30 & 1.50$^{+0.23}_{-0.21}$ & 120.905 $\pm$ 10.946 \\
13.493\,--\,26.915 & 20.204 & 99 & 155.036 $\pm$ 15.582 & 0.78 $\pm$ 0.22 & 0.97$^{+0.13}_{-0.23}$ & 318.766 $\pm$ 32.037 \\
26.915\,--\,53.688 & 40.301 & 49 & 216.076 $\pm$ 30.868 & 0.98 $\pm$ 0.02 & 1.33$^{+0.34}_{-0.38}$ & 384.546 $\pm$ 54.935 \\
53.688\,--\,107.093 & 80.391  & 31 & 385.120 $\pm$ 69.170 & 0.94 $\pm$ 0.06 & 1.62$^{+0.52}_{-0.68}$ & 588.633 $\pm$ 105.722 \\
107.093\,--\,213.623 & 160.358  & 14 & 489.994 $\pm$ 130.956 & 1.00 $\pm$ 0.00 & 2.17$^{+0.89}_{-1.36}$ & 914.329 $\pm$ 244.365 \\
213.623\,--\,426.121 & 319.872 & 9 & 887.429 $\pm$ 295.810 & 1.00 $\pm$ 0.00 & 2.29$^{+1.19}_{-2.71}$ & 2403.157 $\pm$ 801.052 \\
426.121\,--\,850.000 & 638.061 & 5 & 1388.959 $\pm$ 621.161 & 1.00 $\pm$ 0.00  & 1.41$^{+0.20}_{-0.26}$ & 1955.654 $\pm$ 874.595 \\
\hline
\end{tabular}
}
\end{table*}
We used the open-source software {\tt\string AEGEAN}\footnote{\url{https://github.com/PaulHancock/Aegean}} \citep{2012ascl.soft12009H, 2018PASA...35...11H} for simulations. We have injected 400 artificial sources into the residual image extracted from the PyBDSF, following the similar criteria as  \citep{2000A&AS..146...41P, 2016MNRAS.460.2385W}. We have injected 250 point-like sources with source sizes $6.2''- 8.2''$ and 150 extended sources with $\theta_{\text{major}} \geq 15.4''$, while minor axes were assigned using a random axial ratio. The flux densities of sources were in the range $1.7-850$\,mJy and the flux density distribution follows a power law in the form of $dN/dS \propto S^{-1.6}$ \citep{2010A&ARv..18....1D, 2013A&A...549A..55W, 2014MNRAS.440.2791V, 2019MNRAS.490..243C}. The sources were injected randomly throughout the image. We have created 100 stochastic simulations to consider the effects from visibility area and source confusion limitations \citep{2016MNRAS.460.2385W, 2019PASA...36....4F, 2019A&A...622A...4H}. We used PyBDSF, adopting similar parameters to those in section~\ref{sec:catalogue} to extract the original catalogue. The resulting catalogue has been distributed in 9 logarithmic flux bins as raw source counts. The completeness correction has been computed from the ratio of the number of injected sources and the number of recovered sources in each flux bin \citep{2019A&A...622A...4H}. These injection recovery simulations correct for both Eddington bias and resolution bias. The median from 100 simulations for each bin has been used to calculate the completeness correction factor. Figure~\ref{fig:Fig7} shows the median value with their corresponding errors of the 16th and 84th percentiles for completeness correction for each flux density bin in pink pentagons.

\subsection{Differential Source Count} \label{sec:dN/dS}

We have estimated correction factors for FDR and completeness in our catalogue for each logarithmic flux bin. We have multiplied the correction factor by the raw source counts and calculated the Euclidean-normalised differential source count at 183\,MHz for this uGMRT catalogue. We have calculated Poisson errors and considered the effective area per bin to account for noise variation across the image (see the right panel of Figure~\ref{fig:Fig2}). \\
We have compared the corrected differential source count from this work to predicted evolutionary simulated models and other previous observations. We have scaled other source counts at 183\,MHz central frequency with ${\alpha}=-0.7$. We used $S^3-$SKADS \citep{2008MNRAS.388.1335W} and Tiered - Radio Extragalactic Continuum Simulation (T-RECS) II \citep{2023MNRAS.524..993B} simulations for comparison. We have also compared it with observation at various frequencies and fields such as XXL$-$South field at 2.1\,GHz \citep{2018A&A...620A...3B}, ASKAP RACS at 887.5\,MHz \citep{2021PASA...38...58H}, ASKAP EMU GAMA-23 field at 887.5\,MHz \citep{2022MNRAS.512.6104G}, uGMRT GAMA-23 field at 325\,MHz \citep{2026AJ....172..197S}, Bo\"{o}tes field at 400\,MHz  \citep{2023MNRAS.525.5311S}, Lockman Hole field at 325\,MHz \citep{2020MNRAS.495.4071M}, ELAIS-N1 field at 183\,MHz \citep{2025MNRAS.544.3617S}, LoTSS (ELAIS-N1 field) at 150\,MHz \citep{2021A&A...648A...5M} and LOFAR Lockman Hole at 150\,MHz \citep{2016MNRAS.463.2997M}. 
Figure~\ref{fig:Fig8} shows the corrected source counts and their corresponding Poisson errors for the GAMA-23 field obtained using uGMRT at 183\,MHz (represented by maroon circles). The source counts from other observations and simulations are mentioned with their respective markers. Table~\ref{table:dNdS} shows the uGMRT source counts with their corresponding Poisson errors and goes down to 1.7 mJy (${\sim}2$\,mJy). Thus, these counts are limited by the number of sources in the sub-${\mu}$Jy. This limitation is primarily due to the limited on-source observation time of ${\sim}5$ hours. Subsequently, we are bound to dig deeper into the ``drop and bump'' trend by the LoTSS source counts from other deep fields (ELAIS-N1, Lockman Hole and Bo\"{o}tes) at 150\,MHz \citep{2021A&A...648A...5M}. However, the source counts from our uGMRT 183\,MHz catalogue are consistent with other previous observations and simulations.

\section{Conclusion}\label{sec:conclusion}

We present the first observation of the GAMA-23 deep field using the uGMRT Band-2 (120-250\,MHz). The observation is of ${\sim}5$ hours on-source time observed in observing cycle 42. We present the direction-dependent calibration, imaging and catalogue in this work. The final image produced from this observation achieved the off-source RMS noise of 320\,$\mu$Jy beam$^{-1}$. The final image is of field-of-view  $3.5^{\circ}\,{\times}\,3.5^{\circ}$ and resolution of $15.4''$. We have obtained a catalogue of 569 sources using PyBDSF with the dynamic range of $2645$. We have shown comparisons between our catalogue and previous catalogues, such as NVSS at 1400\,MHz, and recent ones, such as ASKAP EMU at 887.5\,MHz and the uGMRT at 325\,MHz for the GAMA-23 field. We estimated the spectral characteristics of the matched sources, which gave the spectral index ${\sim}-0.7$ from the power law $S_\nu \propto \nu^{\alpha}$. We also derive differential source counts to characterise the source population in our catalogue. We estimated the false detection rate and incompleteness in our catalogue and corrected them to obtain the corrected differential source counts. The source counts from this work extend to 1.7\,mJy (${\geq}5{\sigma}$). We compared source counts from this work to simulations and other previous observational surveys. The source counts show consistent behaviour in comparison with both simulations and other surveys. \\
Deep field surveys are important to study the source population in the field. The non-thermal diffuse emission in these deep fields is also crucial for the 21-cm power spectrum from the EoR. The diffuse Galactic synchrotron emission is one of the dominant foregrounds that masks the 21-cm signal. The detailed study of the diffuse Galactic synchrotron emission from this field will be presented in future work.






\section*{Acknowledgements}

RS and AD acknowledge the computing facility used for this work, hosted at the DAASE, IIT Indore, and procured through funding from the Department of Science and Technology, Government of India, sponsored by the DST-FIST grant no. SR/FST/PSII/2021/162 (C) awarded to DAASE, IIT Indore. The authors thank the GMRT staff for providing uGMRT observations. The uGMRT facility is handled by the National Center for Radio Astrophysics (NCRA) of the Tata Institute of Fundamental Research (TIFR). RS acknowledges the support of the UGC-JRF fellowship, granted by UGC. AD acknowledges financial support from the project titled ``Observing the Cosmic Dawn in Multicolor using Next Generation Telescopes'', funded by the Science and Engineering Research Board (SERB), DST, Government of India, through the Core Research Grant No. CRG/2021/004025. This research work used open source package APLpy \footnote{\url{https://pypi.org/project/aplpy/}} and Astropy\footnote{\url{https://www.astropy.org/}} \citep{2013A&A...558A..33A, 2018AJ....156..123A}. 
\vspace{-1em}


\bibliography{sample}{}

@ARTICLE{2017JAI.....641011R,
       author = {{Reddy}, Suda Harshavardhan and {Kudale}, Sanjay and {Gokhale}, Upendra and {Halagalli}, Irappa and {Raskar}, Nilesh and {de}, Kishalay and {Gnanaraj}, Shelton and {Ajith Kumar}, B. and {Gupta}, Yashwant},
        title = "{A Wideband Digital Back-End for the Upgraded GMRT}",
      journal = {Journal of Astronomical Instrumentation},
         year = 2017,
        month = mar,
       volume = {6},
       number = {1},
          eid = {1641011-336},
        pages = {1641011-336},
          doi = {10.1142/S2251171716410117},
       adsurl = {https://ui.adsabs.harvard.edu/abs/2017JAI.....641011R}
}

@ARTICLE{2024MNRAS.528.5346A,
       author = {{An}, Fangxia and {Vaccari}, M. and {Best}, P.~N. and {Ocran}, E.~F. and {Ishwara-Chandra}, C.~H. and {Taylor}, A.~R. and {Leslie}, S.~K. and {R{\"o}ttgering}, H.~J.~A. and {Kondapally}, R. and {Haskell}, Paul and {Collier}, J.~D. and {Bonato}, M.},
        title = "{Radio spectral properties of star-forming galaxies between 150 and 5000 MHz in the ELAIS-N1 field}",
      journal = {\mnras},
         year = 2024,
        month = mar,
       volume = {528},
       number = {3},
        pages = {5346-5363},
          doi = {10.1093/mnras/stae364},
archivePrefix = {arXiv},
       eprint = {2303.06941},
 primaryClass = {astro-ph.GA},
       adsurl = {https://ui.adsabs.harvard.edu/abs/2024MNRAS.528.5346A}
}

@ARTICLE{2010A&ARv..18....1D,
       author = {{de Zotti}, Gianfranco and {Massardi}, Marcella and {Negrello}, Mattia and {Wall}, Jasper},
        title = "{Radio and millimeter continuum surveys and their astrophysical implications}",
      journal = {\aapr},
         year = 2010,
        month = feb,
       volume = {18},
       number = {1-2},
        pages = {1-65},
          doi = {10.1007/s00159-009-0026-0},
archivePrefix = {arXiv},
       eprint = {0908.1896},
 primaryClass = {astro-ph.CO},
       adsurl = {https://ui.adsabs.harvard.edu/abs/2010A&ARv..18....1D}
}

@ARTICLE{1969MNRAS.142....1R,
       author = {{Ringenberg}, R. and {McVittie}, G.~C.},
        title = "{Number counts of extragalactic radio sources}",
      journal = {\mnras},
         year = 1969,
        month = jan,
       volume = {142},
        pages = {1},
          doi = {10.1093/mnras/142.1.1},
       adsurl = {https://ui.adsabs.harvard.edu/abs/1969MNRAS.142....1R}
}

@ARTICLE{2023MNRAS.523.5187W,
       author = {{Wu}, Cossas K.-W. and {Ling}, Chih-Teng and {Goto}, Tomotsugu and {Kim}, Seong Jin and {Hashimoto}, Tetsuya and {Kilerci}, Ece and {Lin}, Yu-Wei and {Wang}, Po-Ya and {Uno}, Yuri and {Ho}, Simon C.-C. and {Hsiao}, Tiger Yu-Yang},
        title = "{Source counts at 7.7-21 {\ensuremath{\mu}}m in CEERS field with JWST}",
      journal = {\mnras},
         year = 2023,
        month = aug,
       volume = {523},
       number = {4},
        pages = {5187-5197},
          doi = {10.1093/mnras/stad1769},
archivePrefix = {arXiv},
       eprint = {2209.01829},
 primaryClass = {astro-ph.GA},
       adsurl = {https://ui.adsabs.harvard.edu/abs/2023MNRAS.523.5187W}
}

@ARTICLE{1998MNRAS.296..839H,
       author = {{Hopkins}, A.~M. and {Mobasher}, B. and {Cram}, L. and {Rowan-Robinson}, M.},
        title = "{The PHOENIX Deep Survey: 1.4-GHz source counts}",
      journal = {\mnras},
         year = 1998,
        month = jun,
       volume = {296},
       number = {4},
        pages = {839-846},
          doi = {10.1046/j.1365-8711.1998.01403.x},
       adsurl = {https://ui.adsabs.harvard.edu/abs/1998MNRAS.296..839H}
}

@ARTICLE{2013MNRAS.436.3759B,
       author = {{Bonzini}, M. and {Padovani}, P. and {Mainieri}, V. and {Kellermann}, K.~I. and {Miller}, N. and {Rosati}, P. and {Tozzi}, P. and {Vattakunnel}, S.},
        title = "{The sub-mJy radio sky in the Extended Chandra Deep Field-South: source population}",
      journal = {\mnras},
         year = 2013,
        month = dec,
       volume = {436},
       number = {4},
        pages = {3759-3771},
          doi = {10.1093/mnras/stt1879},
archivePrefix = {arXiv},
       eprint = {1310.1248},
 primaryClass = {astro-ph.CO},
       adsurl = {https://ui.adsabs.harvard.edu/abs/2013MNRAS.436.3759B}
}

@ARTICLE{2017A&A...602A...2S,
       author = {{Smol{\v{c}}i{\'c}}, V. and {Delvecchio}, I. and {Zamorani}, G. and {Baran}, N. and {Novak}, M. and {Delhaize}, J. and {Schinnerer}, E. and {Berta}, S. and {Bondi}, M. and {Ciliegi}, P. and {Capak}, P. and {Civano}, F. and {Karim}, A. and {Le Fevre}, O. and {Ilbert}, O. and {Laigle}, C. and {Marchesi}, S. and {McCracken}, H.~J. and {Tasca}, L. and {Salvato}, M. and {Vardoulaki}, E.},
        title = "{The VLA-COSMOS 3 GHz Large Project: Multiwavelength counterparts and the composition of the faint radio population}",
      journal = {\aap},
         year = 2017,
        month = jun,
       volume = {602},
          eid = {A2},
        pages = {A2},
          doi = {10.1051/0004-6361/201630223},
archivePrefix = {arXiv},
       eprint = {1703.09719},
 primaryClass = {astro-ph.GA},
       adsurl = {https://ui.adsabs.harvard.edu/abs/2017A&A...602A...2S}
}

@ARTICLE{2010PNAS..107.7184B,
       author = {{Brandt}, W.~N. and {Alexander}, D.~M.},
        title = "{Supermassive black-hole growth over cosmic time: Active galaxy demography, physics, and ecology from Chandra surveys}",
      journal = {Proceedings of the National Academy of Science},
         year = 2010,
        month = apr,
       volume = {107},
        pages = {7184-7189},
          doi = {10.1073/pnas.0914151107},
archivePrefix = {arXiv},
       eprint = {1001.5054},
 primaryClass = {astro-ph.HE},
       adsurl = {https://ui.adsabs.harvard.edu/abs/2010PNAS..107.7184B}
}

@ARTICLE{2024FPP....1100059F,
       author = {{Freundlich}, Jonathan},
        title = "{Star formation across cosmic time}",
      journal = {Fundamental Plasma Physics},
         year = 2024,
        month = sep,
       volume = {11},
          eid = {100059},
        pages = {100059},
          doi = {10.1016/j.fpp.2024.100059},
archivePrefix = {arXiv},
       eprint = {2405.20803},
 primaryClass = {astro-ph.GA},
       adsurl = {https://ui.adsabs.harvard.edu/abs/2024FPP....1100059F}
}

@ARTICLE{2025PASA...42...71H,
       author = {{Hopkins}, Andrew and {Kapinska}, Anna and {Marvil}, Joshua and {Vernstrom}, Tessa and {Collier}, Jordan and {Norris}, Ray and {Gordon}, Yjan and {Duchesne}, Stefan and {Rudnick}, Lawrence and {Gupta}, Nikhel and {Carretti}, Ettore and {Anderson}, Craig and {Dai}, Shi and {G{\"u}rkan}, Gulay and {Parkinson}, David and {Prandoni}, Isabella and {Riggi}, Simone and {Shekhar Saraf}, Chandra and {Ma}, Yik Ki and {Filipovi{\'c}}, Miroslav D. and {Umana}, Grazia and {Bahr-Kalus}, Benedict and {Koribalski}, B{\"a}rbel Silvia and {Lenc}, Emil and {Ingallinera}, Adriano and {Afonso}, Jos{\'e} and {Ahmad}, Adeel and {Ahmed}, Ummee Tania and {Alexander}, Emma and {Andernach}, Heinz and {Asorey}, Jacobo and {Battisti}, Andrew J. and {Bilicki}, Maciej and {Botteon}, Andrea and {Brown}, Michael and {Br{\"u}ggen}, Marcus and {Cowley}, Michael and {Dage}, Kristen and {Hale}, Catherine Laura and {Hardcastle}, Martin and {Kothes}, Roland and {Lazarevi{\'c}}, Sanja and {Lin}, Yen-Ting and {Luken}, Kieran and {Moss}, Jeremy and {Prathap}, P.~K. Jahang and {ur Rahman}, Syed Faisal and {Reiprich}, Thomas and {Riseley}, Christopher and {Salvato}, Mara and {Seymour}, Nicholas and {Shabala}, Stanislav and {Smith}, Daniel and {Vaccari}, Mattia and {van Loon}, Jacco Th. and {Wong}, O. Ivy Ivy and {Zainal Alsaberi}, Rami and {Asher}, Albany and {Ball}, Brianna and {Barbosa}, Davi and {Biava}, Nadia and {Bradley}, Aaron and {Carvajal}, Rodrigo and {Crawford}, Evan J. and {Galvin}, Timothy James and {Huynh}, Minh and {Leahy}, Denis and {Matute}, Israel and {Moss}, Vanessa and {Pappalardo}, Ciro and {Smeaton}, Zachary and {Velovi{\'c}}, Velibor and {Zafar}, Tayyaba},
        title = "{The Evolutionary Map of the Universe: A new radio atlas for the southern hemisphere sky}",
      journal = {\pasa},
         year = 2025,
        month = may,
       volume = {42},
          eid = {e071},
        pages = {e071},
          doi = {10.1017/pasa.2025.10042},
archivePrefix = {arXiv},
       eprint = {2505.08271},
 primaryClass = {astro-ph.GA},
       adsurl = {https://ui.adsabs.harvard.edu/abs/2025PASA...42...71H}
}

@ARTICLE{2016MNRAS.455.3911D,
       author = {{Driver}, Simon P. and {Wright}, Angus H. and {Andrews}, Stephen K. and {Davies}, Luke J. and {Kafle}, Prajwal R. and {Lange}, Rebecca and {Moffett}, Amanda J. and {Mannering}, Elizabeth and {Robotham}, Aaron S.~G. and {Vinsen}, Kevin and {Alpaslan}, Mehmet and {Andrae}, Ellen and {Baldry}, Ivan K. and {Bauer}, Amanda E. and {Bamford}, Steven P. and {Bland-Hawthorn}, Joss and {Bourne}, Nathan and {Brough}, Sarah and {Brown}, Michael J.~I. and {Cluver}, Michelle E. and {Croom}, Scott and {Colless}, Matthew and {Conselice}, Christopher J. and {da Cunha}, Elisabete and {De Propris}, Roberto and {Drinkwater}, Michael and {Dunne}, Loretta and {Eales}, Steve and {Edge}, Alastair and {Frenk}, Carlos and {Graham}, Alister W. and {Grootes}, Meiert and {Holwerda}, Benne W. and {Hopkins}, Andrew M. and {Ibar}, Edo and {van Kampen}, Eelco and {Kelvin}, Lee S. and {Jarrett}, Tom and {Jones}, D. Heath and {Lara-Lopez}, Maritza A. and {Liske}, Jochen and {Lopez-Sanchez}, Angel R. and {Loveday}, Jon and {Maddox}, Steve J. and {Madore}, Barry and {Mahajan}, Smriti and {Meyer}, Martin and {Norberg}, Peder and {Penny}, Samantha J. and {Phillipps}, Steven and {Popescu}, Cristina and {Tuffs}, Richard J. and {Peacock}, John A. and {Pimbblet}, Kevin A. and {Prescott}, Matthew and {Rowlands}, Kate and {Sansom}, Anne E. and {Seibert}, Mark and {Smith}, Matthew W.~L. and {Sutherland}, Will J. and {Taylor}, Edward N. and {Valiante}, Elisabetta and {Vazquez-Mata}, J. Antonio and {Wang}, Lingyu and {Wilkins}, Stephen M. and {Williams}, Richard},
        title = "{Galaxy And Mass Assembly (GAMA): Panchromatic Data Release (far-UV-far-IR) and the low-z energy budget}",
      journal = {\mnras},
         year = 2016,
        month = feb,
       volume = {455},
       number = {4},
        pages = {3911-3942},
          doi = {10.1093/mnras/stv2505},
archivePrefix = {arXiv},
       eprint = {1508.02076},
 primaryClass = {astro-ph.GA},
       adsurl = {https://ui.adsabs.harvard.edu/abs/2016MNRAS.455.3911D}
}

@ARTICLE{2021PASA...38...46N,
       author = {{Norris}, Ray P. and {Marvil}, Joshua and {Collier}, J.~D. and {Kapi{\'n}ska}, Anna D. and {O'Brien}, Andrew N. and {Rudnick}, L. and {Andernach}, Heinz and {Asorey}, Jacobo and {Brown}, Michael J.~I. and {Br{\"u}ggen}, Marcus and {Crawford}, Evan and {English}, Jayanne and {Rahman}, Syed Faisal ur and {Filipovi{\'c}}, Miroslav D. and {Gordon}, Yjan and {G{\"u}rkan}, G{\"u}lay and {Hale}, Catherine and {Hopkins}, Andrew M. and {Huynh}, Minh T. and {HyeongHan}, Kim and {James Jee}, M. and {Koribalski}, B{\"a}rbel S. and {Lenc}, Emil and {Luken}, Kieran and {Parkinson}, David and {Prandoni}, Isabella and {Raja}, Wasim and {Reiprich}, Thomas H. and {Riseley}, Christopher J. and {Shabala}, Stanislav S. and {Sheil}, Jaimie R. and {Vernstrom}, Tessa and {Whiting}, Matthew T. and {Allison}, James R. and {Anderson}, C.~S. and {Ball}, Lewis and {Bell}, Martin and {Bunton}, John and {Galvin}, T.~J. and {Gupta}, Neeraj and {Hotan}, Aidan and {Jacka}, Colin and {Macgregor}, Peter J. and {Mahony}, Elizabeth K. and {Maio}, Umberto and {Moss}, Vanessa and {Pandey-Pommier}, M. and {Voronkov}, Maxim A.},
        title = "{The Evolutionary Map of the Universe pilot survey}",
      journal = {\pasa},
         year = 2021,
        month = sep,
       volume = {38},
          eid = {e046},
        pages = {e046},
          doi = {10.1017/pasa.2021.42},
archivePrefix = {arXiv},
       eprint = {2108.00569},
 primaryClass = {astro-ph.CO},
       adsurl = {https://ui.adsabs.harvard.edu/abs/2021PASA...38...46N}
}

@ARTICLE{2023PASA...40...34D,
       author = {{Duchesne}, S.~W. and {Thomson}, A.~J.~M. and {Pritchard}, J. and {Lenc}, E. and {Moss}, V.~A. and {McConnell}, D. and {Wieringa}, M.~H. and {Whiting}, M.~T. and {Wang}, Z. and {Wang}, Y. and {Rose}, K. and {Raja}, W. and {Murphy}, Tara and {Leung}, J.~K. and {Huynh}, M.~T. and {Hotan}, A.~W. and {Hodgson}, T. and {Heald}, G.~H.},
        title = "{The Rapid ASKAP Continuum Survey IV: continuum imaging at 1367.5 MHz and the first data release of RACS-mid}",
      journal = {\pasa},
         year = 2023,
        month = aug,
       volume = {40},
          eid = {e034},
        pages = {e034},
          doi = {10.1017/pasa.2023.31},
archivePrefix = {arXiv},
       eprint = {2306.07194},
 primaryClass = {astro-ph.IM},
       adsurl = {https://ui.adsabs.harvard.edu/abs/2023PASA...40...34D}
}

@ARTICLE{2026AJ....172..197S,
       author = {{Sagar}, Rashmi and {Datta}, Abhirup and {Mazumder}, Aishrila},
        title = "{Investigating Radio Source Population and Spatial Characteristics of Diffuse Galactic Synchrotron Emission in the GAMA-23 Field with uGMRT Band-3}",
      journal = {\aj},
         year = 2026,
        month = oct,
       volume = {172},
       number = {4},
          eid = {197},
        pages = {197},
          doi = {10.3847/1538-3881/ae9872},
archivePrefix = {arXiv},
       eprint = {2608.05854},
 primaryClass = {astro-ph.GA},
       adsurl = {https://ui.adsabs.harvard.edu/abs/2026AJ....172..197S}
}

@ARTICLE{2015MNRAS.452.1263P,
       author = {{Padovani}, P. and {Bonzini}, M. and {Kellermann}, K.~I. and {Miller}, N. and {Mainieri}, V. and {Tozzi}, P.},
        title = "{Radio-faint AGN: a tale of two populations}",
      journal = {\mnras},
         year = 2015,
        month = sep,
       volume = {452},
       number = {2},
        pages = {1263-1279},
          doi = {10.1093/mnras/stv1375},
archivePrefix = {arXiv},
       eprint = {1506.06554},
 primaryClass = {astro-ph.GA},
       adsurl = {https://ui.adsabs.harvard.edu/abs/2015MNRAS.452.1263P}
}

@ARTICLE{2014PASA...31...41H,
       author = {{Hotan}, A.~W. and {Bunton}, J.~D. and {Harvey-Smith}, L. and {Humphreys}, B. and {Jeffs}, B.~D. and {Shimwell}, T. and {Tuthill}, J. and {Voronkov}, M. and {Allen}, G. and {Amy}, S. and {Ardern}, K. and {Axtens}, P. and {Ball}, L. and {Bannister}, K. and {Barker}, S. and {Bateman}, T. and {Beresford}, R. and {Bock}, D. and {Bolton}, R. and {Bowen}, M. and {Boyle}, B. and {Braun}, R. and {Broadhurst}, S. and {Brodrick}, D. and {Brooks}, K. and {Brothers}, M. and {Brown}, A. and {Cantrall}, C. and {Carrad}, G. and {Chapman}, J. and {Cheng}, W. and {Chippendale}, A. and {Chung}, Y. and {Cooray}, F. and {Cornwell}, T. and {Davis}, E. and {de Souza}, L. and {DeBoer}, D. and {Diamond}, P. and {Edwards}, P. and {Ekers}, R. and {Feain}, I. and {Ferris}, D. and {Forsyth}, R. and {Gough}, R. and {Grancea}, A. and {Gupta}, N. and {Guzman}, J.~C. and {Hampson}, G. and {Haskins}, C. and {Hay}, S. and {Hayman}, D. and {Hoyle}, S. and {Jacka}, C. and {Jackson}, C. and {Jackson}, S. and {Jeganathan}, K. and {Johnston}, S. and {Joseph}, J. and {Kendall}, R. and {Kesteven}, M. and {Kiraly}, D. and {Koribalski}, B. and {Leach}, M. and {Lenc}, E. and {Lensson}, E. and {Li}, L. and {Mackay}, S. and {Macleod}, A. and {Maher}, T. and {Marquarding}, M. and {McClure-Griffiths}, N. and {McConnell}, D. and {Mickle}, S. and {Mirtschin}, P. and {Norris}, R. and {Neuhold}, S. and {Ng}, A. and {O'Sullivan}, J. and {Pathikulangara}, J. and {Pearce}, S. and {Phillips}, C. and {Qiao}, R.~Y. and {Reynolds}, J.~E. and {Rispler}, A. and {Roberts}, P. and {Roxby}, D. and {Schinckel}, A. and {Shaw}, R. and {Shields}, M. and {Storey}, M. and {Sweetnam}, T. and {Troup}, E. and {Turner}, B. and {Tzioumis}, A. and {Westmeier}, T. and {Whiting}, M. and {Wilson}, C. and {Wilson}, T. and {Wormnes}, K. and {Wu}, X.},
        title = "{The Australian Square Kilometre Array Pathfinder: System Architecture and Specifications of the Boolardy Engineering Test Array}",
      journal = {\pasa},
         year = 2014,
        month = nov,
       volume = {31},
          eid = {e041},
        pages = {e041},
          doi = {10.1017/pasa.2014.36},
archivePrefix = {arXiv},
       eprint = {1409.1325},
 primaryClass = {astro-ph.IM},
       adsurl = {https://ui.adsabs.harvard.edu/abs/2014PASA...31...41H}
}

@ARTICLE{2022MNRAS.512.6104G,
       author = {{G{\"u}rkan}, G{\"u}lay and {Prandoni}, I. and {O'Brien}, A. and {Raja}, W. and {Marchetti}, L. and {Vaccari}, M. and {Driver}, S. and {Taylor}, E. and {Franzen}, T. and {Brown}, M.~J.~I. and {Shabala}, S. and {Andernach}, H. and {Hopkins}, A.~M. and {Norris}, R.~P. and {Leahy}, D. and {Bilicki}, M. and {Farajollahi}, H. and {Galvin}, T. and {Heald}, G. and {Koribalski}, B.~S. and {An}, T. and {Warhurst}, K.},
        title = "{Deep ASKAP EMU Survey of the GAMA23 field: properties of radio sources}",
      journal = {\mnras},
         year = 2022,
        month = jun,
       volume = {512},
       number = {4},
        pages = {6104-6121},
          doi = {10.1093/mnras/stac880},
archivePrefix = {arXiv},
       eprint = {2203.14727},
 primaryClass = {astro-ph.GA},
       adsurl = {https://ui.adsabs.harvard.edu/abs/2022MNRAS.512.6104G}
}

@ARTICLE{2021PASA...38...58H,
       author = {{Hale}, Catherine L. and {McConnell}, D. and {Thomson}, A.~J.~M. and {Lenc}, E. and {Heald}, G.~H. and {Hotan}, A.~W. and {Leung}, J.~K. and {Moss}, V.~A. and {Murphy}, T. and {Pritchard}, J. and {Sadler}, E.~M. and {Stewart}, A.~J. and {Whiting}, M.~T.},
        title = "{The Rapid ASKAP Continuum Survey Paper II: First Stokes I Source Catalogue Data Release}",
      journal = {\pasa},
         year = 2021,
        month = nov,
       volume = {38},
          eid = {e058},
        pages = {e058},
          doi = {10.1017/pasa.2021.47},
archivePrefix = {arXiv},
       eprint = {2109.00956},
 primaryClass = {astro-ph.GA},
       adsurl = {https://ui.adsabs.harvard.edu/abs/2021PASA...38...58H}
}

@ARTICLE{2019MNRAS.490..243C,
       author = {{Chakraborty}, Arnab and {Roy}, Nirupam and {Datta}, Abhirup and {Choudhuri}, Samir and {Datta}, Kanan K. and {Dutta}, Prasun and {Bharadwaj}, Somnath and {Intema}, Huib and {Choudhury}, Madhurima and {Pal}, Srijita and {Choudhury}, Tirthankar Roy},
        title = "{Detailed study of ELAIS N1 field with the uGMRT - II. Source properties and spectral variation of foreground power spectrum from 300-500 MHz observations}",
      journal = {\mnras},
         year = 2019,
        month = nov,
       volume = {490},
       number = {1},
        pages = {243-259},
          doi = {10.1093/mnras/stz2533},
archivePrefix = {arXiv},
       eprint = {1908.10380},
 primaryClass = {astro-ph.CO},
       adsurl = {https://ui.adsabs.harvard.edu/abs/2019MNRAS.490..243C}
}

@ARTICLE{2020MNRAS.495.4071M,
       author = {{Mazumder}, Aishrila and {Chakraborty}, Arnab and {Datta}, Abhirup and {Choudhuri}, Samir and {Roy}, Nirupam and {Wadadekar}, Yogesh and {Ishwara-Chandra}, C.~H.},
        title = "{Characterizing EoR foregrounds: a study of the Lockman Hole region at 325 MHz}",
      journal = {\mnras},
         year = 2020,
        month = jul,
       volume = {495},
       number = {4},
        pages = {4071-4084},
          doi = {10.1093/mnras/staa1317},
archivePrefix = {arXiv},
       eprint = {2005.05205},
 primaryClass = {astro-ph.CO},
       adsurl = {https://ui.adsabs.harvard.edu/abs/2020MNRAS.495.4071M}
}

@ARTICLE{2023MNRAS.525.5311S,
       author = {{Sinha}, Akriti and {Datta}, Abhirup},
        title = "{Detailed study of the Bo{\"o}tes field using 300-500 MHz uGMRT observations: source properties and radio-infrared correlations}",
      journal = {\mnras},
         year = 2023,
        month = nov,
       volume = {525},
       number = {4},
        pages = {5311-5327},
          doi = {10.1093/mnras/stad2544},
archivePrefix = {arXiv},
       eprint = {2308.10880},
 primaryClass = {astro-ph.GA},
       adsurl = {https://ui.adsabs.harvard.edu/abs/2023MNRAS.525.5311S}
}

@ARTICLE{2016MNRAS.463.2997M,
       author = {{Mahony}, E.~K. and {Morganti}, R. and {Prandoni}, I. and {van Bemmel}, I.~M. and {Shimwell}, T.~W. and {Brienza}, M. and {Best}, P.~N. and {Br{\"u}ggen}, M. and {Calistro Rivera}, G. and {de Gasperin}, F. and {Hardcastle}, M.~J. and {Harwood}, J.~J. and {Heald}, G. and {Jarvis}, M.~J. and {Mandal}, S. and {Miley}, G.~K. and {Retana-Montenegro}, E. and {R{\"o}ttgering}, H.~J.~A. and {Sabater}, J. and {Tasse}, C. and {van Velzen}, S. and {van Weeren}, R.~J. and {Williams}, W.~L. and {White}, G.~J.},
        title = "{The Lockman Hole project: LOFAR observations and spectral index properties of low-frequency radio sources}",
      journal = {\mnras},
         year = 2016,
        month = dec,
       volume = {463},
       number = {3},
        pages = {2997-3020},
          doi = {10.1093/mnras/stw2225},
archivePrefix = {arXiv},
       eprint = {1609.00537},
 primaryClass = {astro-ph.GA},
       adsurl = {https://ui.adsabs.harvard.edu/abs/2016MNRAS.463.2997M}
}

@ARTICLE{1999A&A...345..380S,
       author = {{Shaver}, P.~A. and {Windhorst}, R.~A. and {Madau}, P. and {de Bruyn}, A.~G.},
        title = "{Can the reionization epoch be detected as a global signature in the cosmic background?}",
      journal = {\aap},
         year = 1999,
        month = may,
       volume = {345},
        pages = {380-390},
          doi = {10.48550/arXiv.astro-ph/9901320},
archivePrefix = {arXiv},
       eprint = {astro-ph/9901320},
 primaryClass = {astro-ph},
       adsurl = {https://ui.adsabs.harvard.edu/abs/1999A&A...345..380S}
}

@INPROCEEDINGS{2003ASSL..285..109G,
       author = {{Greisen}, E.~W.},
        title = "{AIPS, the VLA, and the VLBA}",
    booktitle = {Information Handling in Astronomy - Historical Vistas},
         year = 2003,
       editor = {{Heck}, Andr{\'e}},
       series = {Astrophysics and Space Science Library},
       volume = {285},
        month = mar,
        pages = {109},
          doi = {10.1007/0-306-48080-8_7},
       adsurl = {https://ui.adsabs.harvard.edu/abs/2003ASSL..285..109G}
}

@ARTICLE{2025MNRAS.544.3617S,
       author = {{Sagar}, R. and {Datta}, A. and {Chakraborty}, A. and {Roy}, Nirupam and {Sinha}, A. and {Mazumder}, A. and {Dutta}, Prasun and {Elahi}, Khandakar Md Asif and {Datta}, Kanan K. and {Choudhuri}, Samir and {Bharadwaj}, Somnath and {Pal}, Srijita and {Tripathi}, Anshuman and {Majumdar}, Suman and {Choudhury}, Tirthankar Roy and {Ali}, Sk Saiyad},
        title = "{ELAIS-N1 deep field uGMRT Band-2: constraints on diffuse Galactic synchrotron emission power spectrum}",
      journal = {\mnras},
         year = 2025,
        month = dec,
       volume = {544},
       number = {4},
        pages = {3617-3633},
          doi = {10.1093/mnras/staf1922},
archivePrefix = {arXiv},
       eprint = {2511.02375},
 primaryClass = {astro-ph.HE},
       adsurl = {https://ui.adsabs.harvard.edu/abs/2025MNRAS.544.3617S}
}

@ARTICLE{2016MNRAS.457..730P,
       author = {{Prescott}, Matthew and {Mauch}, T. and {Jarvis}, M.~J. and {McAlpine}, K. and {Smith}, D.~J.~B. and {Fine}, S. and {Johnston}, R. and {Hardcastle}, M.~J. and {Baldry}, I.~K. and {Brough}, S. and {Brown}, M.~J.~I. and {Bremer}, M.~N. and {Driver}, S.~P. and {Hopkins}, A.~M. and {Kelvin}, L.~S. and {Loveday}, J. and {Norberg}, P. and {Obreschkow}, D. and {Sadler}, E.~M.},
        title = "{Galaxy And Mass Assembly (GAMA): the 325 MHz radio luminosity function of AGN and star-forming galaxies}",
      journal = {\mnras},
         year = 2016,
        month = mar,
       volume = {457},
       number = {1},
        pages = {730-744},
          doi = {10.1093/mnras/stv3020},
archivePrefix = {arXiv},
       eprint = {1601.00003},
 primaryClass = {astro-ph.GA},
       adsurl = {https://ui.adsabs.harvard.edu/abs/2016MNRAS.457..730P}
}

@ARTICLE{2016A&ARv..24...13P,
       author = {{Padovani}, Paolo},
        title = "{The faint radio sky: radio astronomy becomes mainstream}",
      journal = {\aapr},
         year = 2016,
        month = sep,
       volume = {24},
       number = {1},
          eid = {13},
        pages = {13},
          doi = {10.1007/s00159-016-0098-6},
archivePrefix = {arXiv},
       eprint = {1609.00499},
 primaryClass = {astro-ph.GA},
       adsurl = {https://ui.adsabs.harvard.edu/abs/2016A&ARv..24...13P}
}

@ARTICLE{2018A&A...620A...3B,
       author = {{Butler}, Andrew and {Huynh}, Minh and {Delhaize}, Jacinta and {Smol{\v{c}}i{\'c}}, Vernesa and {Kapi{\'n}ska}, Anna and {Milakovi{\'c}}, Dinko and {Novak}, Mladen and {Baran}, Nikola and {O'Brien}, Andrew and {Chiappetti}, Lucio and {Desai}, Shantanu and {Fotopoulou}, Sotiria and {Horellou}, Cathy and {Lidman}, Chris and {Pierre}, Marguerite},
        title = "{The XXL Survey. XVIII. ATCA 2.1 GHz radio source catalogue and source counts for the XXL-South field}",
      journal = {\aap},
         year = 2018,
        month = nov,
       volume = {620},
          eid = {A3},
        pages = {A3},
          doi = {10.1051/0004-6361/201630129},
archivePrefix = {arXiv},
       eprint = {1703.10296},
 primaryClass = {astro-ph.GA},
       adsurl = {https://ui.adsabs.harvard.edu/abs/2018A&A...620A...3B}
}

@ARTICLE{2015MNRAS.453.4020F,
       author = {{Franzen}, T.~M.~O. and {Banfield}, J.~K. and {Hales}, C.~A. and {Hopkins}, A. and {Norris}, R.~P. and {Seymour}, N. and {Chow}, K.~E. and {Herzog}, A. and {Huynh}, M.~T. and {Lenc}, E. and {Mao}, M.~Y. and {Middelberg}, E.},
        title = "{ATLAS - I. Third release of 1.4 GHz mosaics and component catalogues}",
      journal = {\mnras},
         year = 2015,
        month = nov,
       volume = {453},
       number = {4},
        pages = {4020-4036},
          doi = {10.1093/mnras/stv1866},
archivePrefix = {arXiv},
       eprint = {1508.03150},
 primaryClass = {astro-ph.GA},
       adsurl = {https://ui.adsabs.harvard.edu/abs/2015MNRAS.453.4020F}
}

@ARTICLE{2020MNRAS.491.1127O,
       author = {{Ocran}, E.~F. and {Taylor}, A.~R. and {Vaccari}, M. and {Ishwara-Chandra}, C.~H. and {Prandoni}, I.},
        title = "{Deep GMRT 610 MHz observations of the ELAIS N1 field: catalogue and source counts}",
      journal = {\mnras},
         year = 2020,
        month = jan,
       volume = {491},
       number = {1},
        pages = {1127-1145},
          doi = {10.1093/mnras/stz2954},
archivePrefix = {arXiv},
       eprint = {1910.08355},
 primaryClass = {astro-ph.GA},
       adsurl = {https://ui.adsabs.harvard.edu/abs/2020MNRAS.491.1127O}
}

@ARTICLE{2011ApJ...739L...1P,
       author = {{Perley}, R.~A. and {Chandler}, C.~J. and {Butler}, B.~J. and {Wrobel}, J.~M.},
        title = "{The Expanded Very Large Array: A New Telescope for New Science}",
      journal = {\apjl},
         year = 2011,
        month = sep,
       volume = {739},
       number = {1},
          eid = {L1},
        pages = {L1},
          doi = {10.1088/2041-8205/739/1/L1},
archivePrefix = {arXiv},
       eprint = {1106.0532},
 primaryClass = {astro-ph.IM},
       adsurl = {https://ui.adsabs.harvard.edu/abs/2011ApJ...739L...1P}
}

@ARTICLE{2018PASA...35...11H,
       author = {{Hancock}, Paul J. and {Trott}, Cathryn M. and {Hurley-Walker}, Natasha},
        title = "{Source Finding in the Era of the SKA (Precursors): Aegean 2.0}",
      journal = {\pasa},
         year = 2018,
        month = mar,
       volume = {35},
          eid = {e011},
        pages = {e011},
          doi = {10.1017/pasa.2018.3},
archivePrefix = {arXiv},
       eprint = {1801.05548},
 primaryClass = {astro-ph.IM},
       adsurl = {https://ui.adsabs.harvard.edu/abs/2018PASA...35...11H}
}

@ARTICLE{2000A&AS..146...41P,
       author = {{Prandoni}, I. and {Gregorini}, L. and {Parma}, P. and {de Ruiter}, H.~R. and {Vettolani}, G. and {Wieringa}, M.~H. and {Ekers}, R.~D.},
        title = "{The ATESP radio survey. II. The source catalogue}",
      journal = {\aaps},
         year = 2000,
        month = oct,
       volume = {146},
        pages = {41-55},
          doi = {10.1051/aas:2000361},
archivePrefix = {arXiv},
       eprint = {astro-ph/0007398},
 primaryClass = {astro-ph},
       adsurl = {https://ui.adsabs.harvard.edu/abs/2000A&AS..146...41P}
}

@ARTICLE{2013A&A...549A..55W,
       author = {{Williams}, W.~L. and {Intema}, H.~T. and {R{\"o}ttgering}, H.~J.~A.},
        title = "{T-RaMiSu: the Two-meter Radio Mini Survey. I. The Bo{\"o}tes Field}",
      journal = {\aap},
         year = 2013,
        month = jan,
       volume = {549},
          eid = {A55},
        pages = {A55},
          doi = {10.1051/0004-6361/201220235},
archivePrefix = {arXiv},
       eprint = {1211.1189},
 primaryClass = {astro-ph.CO},
       adsurl = {https://ui.adsabs.harvard.edu/abs/2013A&A...549A..55W}
}

@ARTICLE{2008MNRAS.388.1335W,
       author = {{Wilman}, R.~J. and {Miller}, L. and {Jarvis}, M.~J. and {Mauch}, T. and {Levrier}, F. and {Abdalla}, F.~B. and {Rawlings}, S. and {Kl{\"o}ckner}, H. -R. and {Obreschkow}, D. and {Olteanu}, D. and {Young}, S.},
        title = "{A semi-empirical simulation of the extragalactic radio continuum sky for next generation radio telescopes}",
      journal = {\mnras},
         year = 2008,
        month = aug,
       volume = {388},
       number = {3},
        pages = {1335-1348},
          doi = {10.1111/j.1365-2966.2008.13486.x},
archivePrefix = {arXiv},
       eprint = {0805.3413},
 primaryClass = {astro-ph},
       adsurl = {https://ui.adsabs.harvard.edu/abs/2008MNRAS.388.1335W}
}

@ARTICLE{2021A&A...648A...5M,
       author = {{Mandal}, S. and {Prandoni}, I. and {Hardcastle}, M.~J. and {Shimwell}, T.~W. and {Intema}, H.~T. and {Tasse}, C. and {van Weeren}, R.~J. and {Algera}, H. and {Emig}, K.~L. and {R{\"o}ttgering}, H.~J.~A. and {Schwarz}, D.~J. and {Siewert}, T.~M. and {Best}, P.~N. and {Bonato}, M. and {Bondi}, M. and {Jarvis}, M.~J. and {Kondapally}, R. and {Leslie}, S.~K. and {Mahatma}, V.~H. and {Sabater}, J. and {Retana-Montenegro}, E. and {Williams}, W.~L.},
        title = "{Extremely deep 150 MHz source counts from the LoTSS Deep Fields}",
      journal = {\aap},
         year = 2021,
        month = apr,
       volume = {648},
          eid = {A5},
        pages = {A5},
          doi = {10.1051/0004-6361/202039998},
archivePrefix = {arXiv},
       eprint = {2011.08829},
 primaryClass = {astro-ph.GA},
       adsurl = {https://ui.adsabs.harvard.edu/abs/2021A&A...648A...5M}
}

@ARTICLE{2013A&A...558A..33A,
       author = {{Astropy Collaboration} and {Robitaille}, Thomas P. and {Tollerud}, Erik J. and {Greenfield}, Perry and {Droettboom}, Michael and {Bray}, Erik and {Aldcroft}, Tom and {Davis}, Matt and {Ginsburg}, Adam and {Price-Whelan}, Adrian M. and {Kerzendorf}, Wolfgang E. and {Conley}, Alexander and {Crighton}, Neil and {Barbary}, Kyle and {Muna}, Demitri and {Ferguson}, Henry and {Grollier}, Fr{\'e}d{\'e}ric and {Parikh}, Madhura M. and {Nair}, Prasanth H. and {Unther}, Hans M. and {Deil}, Christoph and {Woillez}, Julien and {Conseil}, Simon and {Kramer}, Roban and {Turner}, James E.~H. and {Singer}, Leo and {Fox}, Ryan and {Weaver}, Benjamin A. and {Zabalza}, Victor and {Edwards}, Zachary I. and {Azalee Bostroem}, K. and {Burke}, D.~J. and {Casey}, Andrew R. and {Crawford}, Steven M. and {Dencheva}, Nadia and {Ely}, Justin and {Jenness}, Tim and {Labrie}, Kathleen and {Lim}, Pey Lian and {Pierfederici}, Francesco and {Pontzen}, Andrew and {Ptak}, Andy and {Refsdal}, Brian and {Servillat}, Mathieu and {Streicher}, Ole},
        title = "{Astropy: A community Python package for astronomy}",
      journal = {\aap},
         year = 2013,
        month = oct,
       volume = {558},
          eid = {A33},
        pages = {A33},
          doi = {10.1051/0004-6361/201322068},
archivePrefix = {arXiv},
       eprint = {1307.6212},
 primaryClass = {astro-ph.IM},
       adsurl = {https://ui.adsabs.harvard.edu/abs/2013A&A...558A..33A}
}

@ARTICLE{1998AJ....115.1693C,
       author = {{Condon}, J.~J. and {Cotton}, W.~D. and {Greisen}, E.~W. and {Yin}, Q.~F. and {Perley}, R.~A. and {Taylor}, G.~B. and {Broderick}, J.~J.},
        title = "{The NRAO VLA Sky Survey}",
      journal = {\aj},
         year = 1998,
        month = may,
       volume = {115},
       number = {5},
        pages = {1693-1716},
          doi = {10.1086/300337},
       adsurl = {https://ui.adsabs.harvard.edu/abs/1998AJ....115.1693C}
}

@ARTICLE{2018AJ....156..123A,
       author = {{Astropy Collaboration} and {Price-Whelan}, A.~M. and {Sip{\H{o}}cz}, B.~M. and {G{\"u}nther}, H.~M. and {Lim}, P.~L. and {Crawford}, S.~M. and {Conseil}, S. and {Shupe}, D.~L. and {Craig}, M.~W. and {Dencheva}, N. and {Ginsburg}, A. and {VanderPlas}, J.~T. and {Bradley}, L.~D. and {P{\'e}rez-Su{\'a}rez}, D. and {de Val-Borro}, M. and {Aldcroft}, T.~L. and {Cruz}, K.~L. and {Robitaille}, T.~P. and {Tollerud}, E.~J. and {Ardelean}, C. and {Babej}, T. and {Bach}, Y.~P. and {Bachetti}, M. and {Bakanov}, A.~V. and {Bamford}, S.~P. and {Barentsen}, G. and {Barmby}, P. and {Baumbach}, A. and {Berry}, K.~L. and {Biscani}, F. and {Boquien}, M. and {Bostroem}, K.~A. and {Bouma}, L.~G. and {Brammer}, G.~B. and {Bray}, E.~M. and {Breytenbach}, H. and {Buddelmeijer}, H. and {Burke}, D.~J. and {Calderone}, G. and {Cano Rodr{\'\i}guez}, J.~L. and {Cara}, M. and {Cardoso}, J.~V.~M. and {Cheedella}, S. and {Copin}, Y. and {Corrales}, L. and {Crichton}, D. and {D'Avella}, D. and {Deil}, C. and {Depagne}, {\'E}. and {Dietrich}, J.~P. and {Donath}, A. and {Droettboom}, M. and {Earl}, N. and {Erben}, T. and {Fabbro}, S. and {Ferreira}, L.~A. and {Finethy}, T. and {Fox}, R.~T. and {Garrison}, L.~H. and {Gibbons}, S.~L.~J. and {Goldstein}, D.~A. and {Gommers}, R. and {Greco}, J.~P. and {Greenfield}, P. and {Groener}, A.~M. and {Grollier}, F. and {Hagen}, A. and {Hirst}, P. and {Homeier}, D. and {Horton}, A.~J. and {Hosseinzadeh}, G. and {Hu}, L. and {Hunkeler}, J.~S. and {Ivezi{\'c}}, {\v{Z}}. and {Jain}, A. and {Jenness}, T. and {Kanarek}, G. and {Kendrew}, S. and {Kern}, N.~S. and {Kerzendorf}, W.~E. and {Khvalko}, A. and {King}, J. and {Kirkby}, D. and {Kulkarni}, A.~M. and {Kumar}, A. and {Lee}, A. and {Lenz}, D. and {Littlefair}, S.~P. and {Ma}, Z. and {Macleod}, D.~M. and {Mastropietro}, M. and {McCully}, C. and {Montagnac}, S. and {Morris}, B.~M. and {Mueller}, M. and {Mumford}, S.~J. and {Muna}, D. and {Murphy}, N.~A. and {Nelson}, S. and {Nguyen}, G.~H. and {Ninan}, J.~P. and {N{\"o}the}, M. and {Ogaz}, S. and {Oh}, S. and {Parejko}, J.~K. and {Parley}, N. and {Pascual}, S. and {Patil}, R. and {Patil}, A.~A. and {Plunkett}, A.~L. and {Prochaska}, J.~X. and {Rastogi}, T. and {Reddy Janga}, V. and {Sabater}, J. and {Sakurikar}, P. and {Seifert}, M. and {Sherbert}, L.~E. and {Sherwood-Taylor}, H. and {Shih}, A.~Y. and {Sick}, J. and {Silbiger}, M.~T. and {Singanamalla}, S. and {Singer}, L.~P. and {Sladen}, P.~H. and {Sooley}, K.~A. and {Sornarajah}, S. and {Streicher}, O. and {Teuben}, P. and {Thomas}, S.~W. and {Tremblay}, G.~R. and {Turner}, J.~E.~H. and {Terr{\'o}n}, V. and {van Kerkwijk}, M.~H. and {de la Vega}, A. and {Watkins}, L.~L. and {Weaver}, B.~A. and {Whitmore}, J.~B. and {Woillez}, J. and {Zabalza}, V. and {Astropy Contributors}},
        title = "{The Astropy Project: Building an Open-science Project and Status of the v2.0 Core Package}",
      journal = {\aj},
         year = 2018,
        month = sep,
       volume = {156},
       number = {3},
          eid = {123},
        pages = {123},
          doi = {10.3847/1538-3881/aabc4f},
archivePrefix = {arXiv},
       eprint = {1801.02634},
 primaryClass = {astro-ph.IM},
       adsurl = {https://ui.adsabs.harvard.edu/abs/2018AJ....156..123A}
}

@ARTICLE{2016MNRAS.460.2385W,
       author = {{Williams}, W.~L. and {van Weeren}, R.~J. and {R{\"o}ttgering}, H.~J.~A. and {Best}, P. and {Dijkema}, T.~J. and {de Gasperin}, F. and {Hardcastle}, M.~J. and {Heald}, G. and {Prandoni}, I. and {Sabater}, J. and {Shimwell}, T.~W. and {Tasse}, C. and {van Bemmel}, I.~M. and {Br{\"u}ggen}, M. and {Brunetti}, G. and {Conway}, J.~E. and {En{\ss}lin}, T. and {Engels}, D. and {Falcke}, H. and {Ferrari}, C. and {Haverkorn}, M. and {Jackson}, N. and {Jarvis}, M.~J. and {Kapi{\'n}ska}, A.~D. and {Mahony}, E.~K. and {Miley}, G.~K. and {Morabito}, L.~K. and {Morganti}, R. and {Orr{\'u}}, E. and {Retana-Montenegro}, E. and {Sridhar}, S.~S. and {Toribio}, M.~C. and {White}, G.~J. and {Wise}, M.~W. and {Zwart}, J.~T.~L.},
        title = "{LOFAR 150-MHz observations of the Bo{\"o}tes field: catalogue and source counts}",
      journal = {\mnras},
         year = 2016,
        month = aug,
       volume = {460},
       number = {3},
        pages = {2385-2412},
          doi = {10.1093/mnras/stw1056},
archivePrefix = {arXiv},
       eprint = {1605.01531},
 primaryClass = {astro-ph.CO},
       adsurl = {https://ui.adsabs.harvard.edu/abs/2016MNRAS.460.2385W}
}

@ARTICLE{2023MNRAS.523.1729B,
       author = {{Best}, P.~N. and {Kondapally}, R. and {Williams}, W.~L. and {Cochrane}, R.~K. and {Duncan}, K.~J. and {Hale}, C.~L. and {Haskell}, P. and {Ma{\l}ek}, K. and {McCheyne}, I. and {Smith}, D.~J.~B. and {Wang}, L. and {Botteon}, A. and {Bonato}, M. and {Bondi}, M. and {Calistro Rivera}, G. and {Gao}, F. and {G{\"u}rkan}, G. and {Hardcastle}, M.~J. and {Jarvis}, M.~J. and {Mingo}, B. and {Miraghaei}, H. and {Morabito}, L.~K. and {Nisbet}, D. and {Prandoni}, I. and {R{\"o}ttgering}, H.~J.~A. and {Sabater}, J. and {Shimwell}, T. and {Tasse}, C. and {van Weeren}, R.},
        title = "{The LOFAR Two-metre Sky Survey: Deep Fields data release 1. V. Survey description, source classifications, and host galaxy properties}",
      journal = {\mnras},
         year = 2023,
        month = aug,
       volume = {523},
       number = {2},
        pages = {1729-1755},
          doi = {10.1093/mnras/stad1308},
archivePrefix = {arXiv},
       eprint = {2305.05782},
 primaryClass = {astro-ph.GA},
       adsurl = {https://ui.adsabs.harvard.edu/abs/2023MNRAS.523.1729B}
}

@ARTICLE{2006A&A...457..517P,
       author = {{Prandoni}, I. and {Parma}, P. and {Wieringa}, M.~H. and {de Ruiter}, H.~R. and {Gregorini}, L. and {Mignano}, A. and {Vettolani}, G. and {Ekers}, R.~D.},
        title = "{The ATESP 5 GHz radio survey. I. Source counts and spectral index properties of the faint radio population}",
      journal = {\aap},
         year = 2006,
        month = oct,
       volume = {457},
       number = {2},
        pages = {517-529},
          doi = {10.1051/0004-6361:20054273},
archivePrefix = {arXiv},
       eprint = {astro-ph/0607141},
 primaryClass = {astro-ph},
       adsurl = {https://ui.adsabs.harvard.edu/abs/2006A&A...457..517P}
}

@ARTICLE{2004MNRAS.355.1053D,
       author = {{Di Matteo}, Tiziana and {Ciardi}, Benedetta and {Miniati}, Francesco},
        title = "{The 21-cm emission from the reionization epoch: extended and point source foregrounds}",
      journal = {\mnras},
         year = 2004,
        month = dec,
       volume = {355},
       number = {4},
        pages = {1053-1065},
          doi = {10.1111/j.1365-2966.2004.08443.x},
archivePrefix = {arXiv},
       eprint = {astro-ph/0402322},
 primaryClass = {astro-ph},
       adsurl = {https://ui.adsabs.harvard.edu/abs/2004MNRAS.355.1053D}
}

@ARTICLE{2017CSci..113..707G,
       author = {{Gupta}, Y. and {Ajithkumar}, B. and {Kale}, H.~S. and {Nayak}, S. and {Sabhapathy}, S. and {Sureshkumar}, S. and {Swami}, R.~V. and {Chengalur}, J.~N. and {Ghosh}, S.~K. and {Ishwara-Chandra}, C.~H. and {Joshi}, B.~C. and {Kanekar}, N. and {Lal}, D.~V. and {Roy}, S.},
        title = "{The upgraded GMRT: opening new windows on the radio Universe}",
      journal = {Current Science},
         year = 2017,
        month = aug,
       volume = {113},
       number = {4},
        pages = {707-714},
          doi = {10.18520/cs/v113/i04/707-714},
       adsurl = {https://ui.adsabs.harvard.edu/abs/2017CSci..113..707G}
}

@ARTICLE{2013PASA...30...31B,
       author = {{Bowman}, Judd D. and {Cairns}, Iver and {Kaplan}, David L. and {Murphy}, Tara and {Oberoi}, Divya and {Staveley-Smith}, Lister and {Arcus}, Wayne and {Barnes}, David G. and {Bernardi}, Gianni and {Briggs}, Frank H. and {Brown}, Shea and {Bunton}, John D. and {Burgasser}, Adam J. and {Cappallo}, Roger J. and {Chatterjee}, Shami and {Corey}, Brian E. and {Coster}, Anthea and {Deshpande}, Avinash and {deSouza}, Ludi and {Emrich}, David and {Erickson}, Philip and {Goeke}, Robert F. and {Gaensler}, B.~M. and {Greenhill}, Lincoln J. and {Harvey-Smith}, Lisa and {Hazelton}, Bryna J. and {Herne}, David and {Hewitt}, Jacqueline N. and {Johnston-Hollitt}, Melanie and {Kasper}, Justin C. and {Kincaid}, Barton B. and {Koenig}, Ronald and {Kratzenberg}, Eric and {Lonsdale}, Colin J. and {Lynch}, Mervyn J. and {Matthews}, Lynn D. and {McWhirter}, S. Russell and {Mitchell}, Daniel A. and {Morales}, Miguel F. and {Morgan}, Edward H. and {Ord}, Stephen M. and {Pathikulangara}, Joseph and {Prabu}, Thiagaraj and {Remillard}, Ronald A. and {Robishaw}, Timothy and {Rogers}, Alan E.~E. and {Roshi}, Anish A. and {Salah}, Joseph E. and {Sault}, Robert J. and {Shankar}, N. Udaya and {Srivani}, K.~S. and {Stevens}, Jamie B. and {Subrahmanyan}, Ravi and {Tingay}, Steven J. and {Wayth}, Randall B. and {Waterson}, Mark and {Webster}, Rachel L. and {Whitney}, Alan R. and {Williams}, Andrew J. and {Williams}, Christopher L. and {Wyithe}, J. Stuart B.},
        title = "{Science with the Murchison Widefield Array}",
      journal = {\pasa},
         year = 2013,
        month = apr,
       volume = {30},
          eid = {e031},
        pages = {e031},
          doi = {10.1017/pas.2013.009},
archivePrefix = {arXiv},
       eprint = {1212.5151},
 primaryClass = {astro-ph.IM},
       adsurl = {https://ui.adsabs.harvard.edu/abs/2013PASA...30...31B}
}
\bibliographystyle{apj}
\end{document}